\documentclass[11pt]{article}
\usepackage[DIV10]{typearea}

\usepackage[latin1]{inputenc}
\usepackage[english]{babel}

\usepackage{amsfonts}
\usepackage{amsmath}    
\usepackage{amssymb}
\usepackage{amsthm}
\usepackage{tensor}
\usepackage{upgreek}
\usepackage{bbm}        
\usepackage{braket}
\usepackage[pdftex]{color}
\usepackage{wasysym}
\usepackage{slashed}
\usepackage{multirow, booktabs}
\usepackage{graphicx}
\usepackage{afterpage}
\usepackage{exscale}
\usepackage{subfigure}        

\usepackage[pdftex,colorlinks,pdfpagelabels]{hyperref}
\usepackage[figure,table]{hypcap} 
\hypersetup{
   bookmarksnumbered,
   pdfstartview={FitV},
   pdfpagemode={UseOutlines},
   pdfauthor={Jasper Hasenkamp, Joern Kersten},
   pdftitle={Dark radiation from particle decay: cosmological constraints and opportunities},
   pdfsubject={scientific publication}
   ,citecolor={blue},
   linkcolor={blue},
   urlcolor={blue},
   filecolor={blue}
} 

\def\a{\alpha}

\def\d{\delta}
\def\e{\epsilon}

\def\g{\gamma}

\def\k{\kappa}
\def\l{\lambda}
\def\m{\mu}
\def\n{\nu}
\def\o{\omega}

\def\r{\rho}
\def\s{\sigma}
\def\t{\tau}

\def\D{\Delta}

\def\G{\Gamma}

\def\O{\Omega}

\newcommand{\eV}{\text{ eV}}
\newcommand{\keV}{\text{ keV}}
\newcommand{\MeV}{\text{ MeV}}
\newcommand{\GeV}{\text{ GeV}}
\newcommand{\TeV}{\text{ TeV}}
\newcommand{\Mpc}{\text{ Mpc}}

\newcommand{\seconds}{\text{ s}}

\newcommand{\mgrav}{m_{3/2}}

\newcommand{\msax}{m_\text{sax}}

\newcommand{\mplanck}{\ensuremath{M_{\text{pl}}}}

\newcommand{\omegab}{\O_{\text{b}}h^2}
\newcommand{\omegadm}{\O_{\text{dm}} h^2}
\newcommand{\rhocrit}{\r_{\text{c}}}
\newcommand{\teq}{{t_\text{eq}}}
\newcommand{\Teq}{T_\text{eq}}
\newcommand{\zeq}{z_\text{eq}}

\newcommand{\Neff}{N_\text{eff}}

\newcommand{\gast}{g_\ast}
\newcommand{\gastd}{g_\ast^\text{d}}
\newcommand{\gastnr}{g_\ast^\text{nr}}
\newcommand{\gastt}{g_\ast^0}
\newcommand{\gasteq}{g_\ast^\text{eq}}

\newcommand{\gasts}{g_{\ast s}}
\newcommand{\gastsd}{g_{\ast s}^\text{d}}
\newcommand{\gastsnr}{g_{\ast s}^\text{nr}}
\newcommand{\gastst}{g_{\ast s}^0}

\newcommand{\gdr}{g_\text{dr}}
\newcommand{\gdrobs}{g_\text{dr}^\text{obs}}

\newcommand{\Tlow}{T_\text{low}}
\newcommand{\rhodr}{\rho_\text{dr}}
\newcommand{\rhorad}{\rho_\text{rad}}

\newcommand{\Tdec}{T_\text{d}}
\newcommand{\adec}{a_\text{d}}
\newcommand{\Tnr}{T^\text{nr}}

\newcommand{\tnrmin}{(t_2^\text{nr})_\text{min}}
\newcommand{\bmax}{b_\text{max}}
\newcommand{\taumax}{t_\text{cmb}}

\begin{document}
 
\date{\mbox{ }}

\title{ 
\bf Dark radiation from particle decay: \\
cosmological constraints and opportunities
\\[8mm]}
\author{Jasper Hasenkamp and J\"{o}rn Kersten\\[2mm]
{\small\it II.~Institute for Theoretical Physics, University of Hamburg,
22761 Hamburg, Germany}\\
{\small\tt Jasper.Hasenkamp@desy.de, Joern.Kersten@desy.de}
}
\maketitle

\thispagestyle{empty}

\vspace{1cm}

\begin{abstract}
\noindent
We study particle decay as the origin of dark radiation.
After elaborating general properties and useful parametrisations we provide model-independent and easy-to-use constraints from nucleosynthesis, the cosmic microwave background and structure formation.
Bounds on branching ratios and mass hierarchies depend in a unique way on the time of decay.
We demonstrate their power to exclude well-motivated scenarios taking the example of the lightest ordinary sparticle decaying into the gravitino.
We point out signatures and opportunities in cosmological observations and structure formation.
For example, if there are two dark decay modes, dark radiation and the
observed dark matter with adjustable free-streaming can originate from
the same decaying particle, solving small-scale problems of structure formation.
Hot dark matter mimicking a neutrino mass scale as deduced from
cosmological observations can arise and possibly be distinguished after
a discovery.
Our results can be used as a guideline for model building.
\end{abstract}

\newpage

\section{Introduction}
New cosmological probes measure the amount of radiation in the Universe at different epochs of its evolution with a crucial increase in precision. 
One strength of the standard cosmological model, which is based upon general relativity and the Standard Model (SM) of particle physics amended by ``invisible'' components known as dark matter and dark energy, is the precise prediction of the amount of radiation.
As usual radiation refers to any relativistic particle. Its amount is often given in terms of the parameter $\Neff$. 
Within the first 20 minutes light nuclei like helium were formed  during the process of big bang nucleosynthesis (BBN)  
 as observed today. 
At such early times the Universe was dominated by radiation. Since nucleosynthesis depends on the expansion rate, BBN is sensitive to the amount of radiation.
There is still a controversy between different groups, some favouring the prediction and others an increased amount~\cite{Izotov:2010ca,Aver:2010wq,Mangano:2011ar}.
Since the main uncertainty stems from the determination of the relic helium abundance from observations, we can expect improvements in the foreseeable future.
Observations of the cosmic microwave background (CMB) constrain the amount of radiation in the Universe in an epoch lasting from some thousand years till photons decouple $10^5$ years later.
Since the first determination of the radiation content of the Universe from the CMB roughly ten years ago~\cite{Bowen:2001in} and also in current measurements by the South Pole Telescope~\cite{Keisler:2011aw} and the Atacama Cosmology Telescope~\cite{Dunkley:2010ge}, mean values are larger than the prediction.
The observed suppression of the CMB power spectrum at larger multipoles would be due to increased 
Silk damping~\cite{Hou:2011ec}. 
Non-Gaussianities could provide further insights~\cite{Kawakami:2012ke}.
Additional radiation also eliminates tension between cosmological data and measurements of today's expansion rate~\cite{Mehta:2012hh,Calabrese:2012vf}.
Increased mean values are found as well in extended analyses including additional cosmological data~\cite{Bashinsky:2003tk,Hamann:2007pi,GonzalezGarcia:2010un,Calabrese:2011hg,GonzalezMorales:2011ty,Moresco:2012by,Joudaki:2012fx,Archidiacono:2012gv}, although the results for the statistical significance of
this increase vary.
More importantly, due to the increase in precision~\cite{Perotto:2006rj,Hamann:2007sb} the Planck satellite, which finished data taking already, could turn these hints into a 3-$\sigma$ to 5-$\s$ discovery, if current mean values are accurate. 
Our understanding of the third component in the Universe --besides matter and vacuum energy-- would be proven incomplete, too.
Since such additional, ``invisible'' radiation cannot arise from the SM and its nature is unknown, it has been dubbed dark radiation.

In this work we study particle decay as the origin of dark radiation.
Additional radiation has been studied mainly in connection with the
possible existence of additional neutrino species beyond the known
three~\cite{Hamann:2010bk,Abazajian:2012ys,Mirizzi:2012we,SolagurenBeascoa:2012cz,Anchordoqui:2012qu} or other relativistic species~\cite{Jaeckel:2008fi,Nakayama:2010vs,Feng:2011uf}.
The amount of additional radiation is then generically discrete, fixed by the spin and number of internal degrees of
freedom of the particles. 
The species are relativistic during BBN and may or may not still be relativistic around photon decoupling.
Hence, they always lead to an increase in radiation during BBN and
typically to the same increase during CMB times.
This is an appealing prediction, in particular, because
past observations have been too imprecise to find a difference between BBN and CMB determinations.
In contrast, dark radiation from particle decay can originate before~\cite{Chang:1996ih,Hasenkamp:2011xh,Jeong:2012np,Graf:2012hb,Cicoli:2012aq,Higaki:2012ar}, during or after BBN~\cite{Ichikawa:2007jv,Fischler:2010xz,Hasenkamp:2011em,Fuller:2011qy,Menestrina:2011mz,Hooper:2011aj,Bjaelde:2012wi,Choi:2012zna} and can become non-relativistic before, during or after photon decoupling. Any observed increase in $\Neff$ can be explained.
These are qualitative differences to the case of a relativistic species that become distinguishable given the new observational precision.
Even though current observations do not allow for any conclusion~\cite{Hou:2011ec,GonzalezGarcia:2012yq}, the most interesting finding would be an increase in radiation after BBN\@. This would exclude additional relativistic species leaving late particle decay as the most attractive possibility.\footnote{
One alternative is the late annihilation of additional species~\cite{Boehm:2012gr}.
}
Interestingly, the time of decay can be probed in the inflationary gravitational wave background~\cite{Jinno:2012xb}.

We exploit the fact that the energy density of the decaying particle is fixed by the observed amount of dark radiation to determine model-independent upper bounds on several branching ratios of the decaying particle from BBN, spectral distortions in the CMB and the ionisation history of the Universe. We point out an opportunity to solve the cosmological lithium problems~\cite{Jedamzik:1999di} and the discovery potential of a future CMB polarimeter for the considered decay.
A decay before BBN could mimic a cosmology with additional relativistic species. More importantly,
there is a plethora of new cosmologies.
We elaborate constraints and opportunities relating to heavier decay products in structure formation.
They may form dark radiation, but they do not need to act as radiation at all. 
If they form the observed dark matter, two of three dark components would originate from the same decay.
If they are not cold, their free-streaming might resolve the missing satellites problem~\cite{Klypin:1999uc,Moore:1999nt}.
While lighter decay products act as dark radiation, heavier ones might mimic the neutrino mass scale as deduced from cosmological observations.
Cluster abundances seem to favour additional radiation together with a finite neutrino mass scale~\cite{Benson:2011ut}.

In the next section we study the simplest case allowing for exactly one dark decay mode.
We will use our findings in Sec.~\ref{sec:tddms} to explore which opportunities open up in structure formation, if there are two dark decay modes.
Sec.~\ref{sec:brconstraints} is devoted to general constraints and
opportunities from BBN and the CMB\@.
We summarise and conclude in Sec.~\ref{sec:conclusions}.
In the appendix we provide an analytic treatment of the exponential decay law in an expanding universe.

\section{One dark decay mode}
\label{sec:oddms}
\begin{figure}[ht]
 \centering
   \includegraphics[width=0.9 \textwidth]{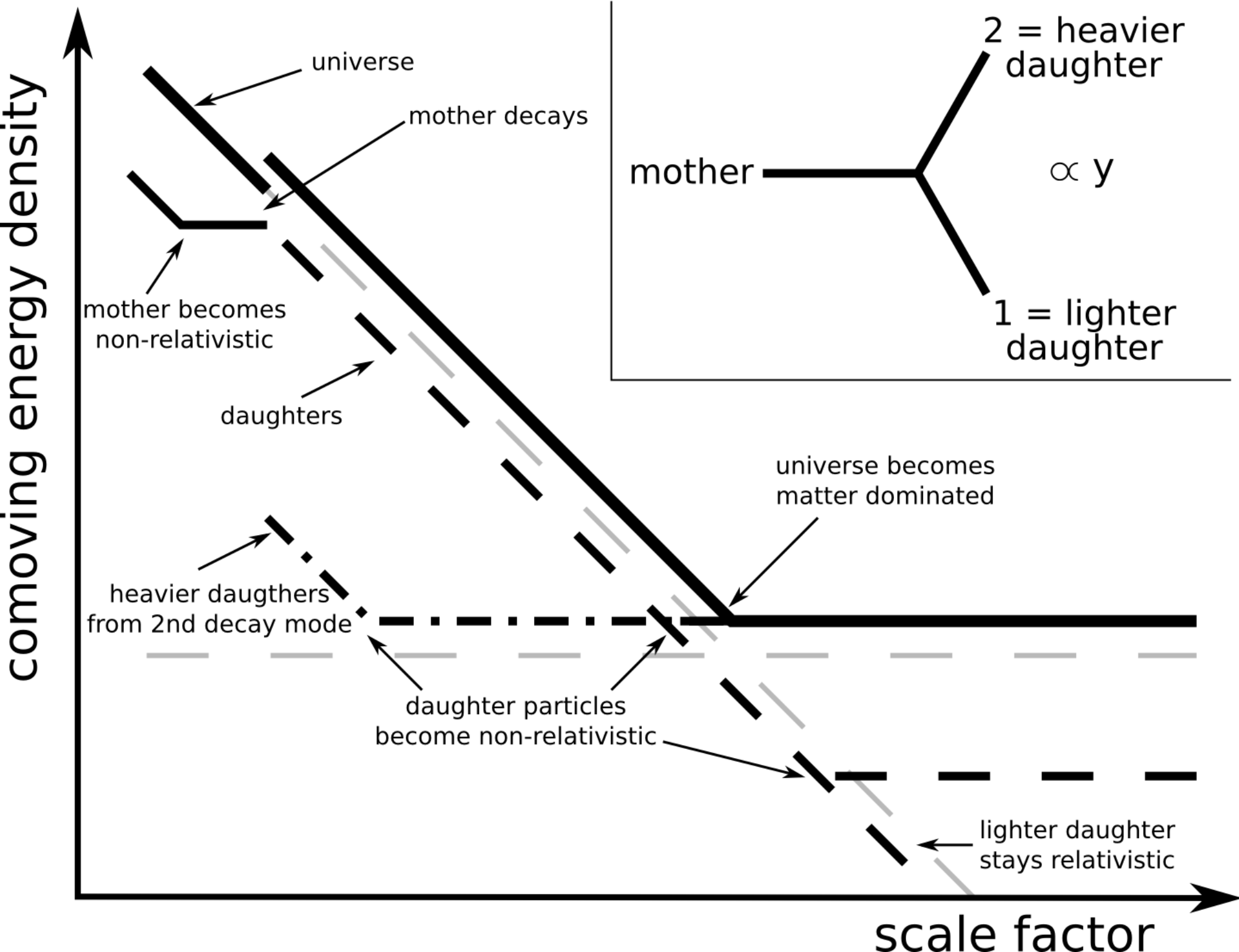}
\caption[energydensities]{Behaviour of comoving energy densities $\r a^3$ in an expanding universe with dark radiation from particle decay. 
The full-logarithmic figure is illustrative and not exact. 
Upper right corner: Nomenclature for the considered two-body decay.
}
\label{fig:energydensities}
 \end{figure}
In this section we study the origin of dark radiation from (one) two-body decay of a non-relativistic particle.
Such a decay is drawn in the upper right corner of Fig.~\ref{fig:energydensities}, where we indicate some nomenclature.
The decaying particle (mother) decays into non-identical particles, where one is heavier (heavier daughter) and the other one lighter (lighter daughter).
In Fig.~\ref{fig:energydensities} we illustrate the behaviour of certain energy densities in a cosmology with dark radiation from particle decay. First of all, the plethora of possible cosmologies cannot be shown in only one figure. So we illustrate a few typical and interesting scenarios.
At some early time corresponding to a small scale factor 
the Universe is dominated by radiation and the mother is relativistic, so their energy densities scale equally.
At some time the mother becomes non-relativistic and from then on its energy density $\r \propto a^{-3}$ grows relative to the radiation energy density $\rhorad \propto a^{-4}$ as the Universe expands, where $a$ denotes the growing scale factor.
Thus stable or very long-lived matter generically comes to dominate the Universe.
Actually, this fact gives rise to various cosmological problems with the well-known gravitino problem as prime example.
These problems may turn out as fortunes, if they give rise to the desired dark radiation as in~\cite{Hasenkamp:2011em}.
Since only the relative behaviour matters, comoving energy densities $\r a^3$ are drawn.
When the mother decays, its energy density is converted into the energy density of its relativistic daughters.
The radiation content of the Universe is increased. 
Of course, there are various possible production mechanisms for the mother in the early universe. It is crucial only that its energy density is within a certain range at its decay, see below.
The energy density of the daughters scales as radiation till they become
non-relativistic. They may still be relativistic today or, particularly
the heavier one, may have become non-relativistic earlier and thus could possibly form the observed dark matter or some hot dark matter component.
For comparison, the energy densities of radiation and matter in standard cosmology are shown as grey dashed curves.

If the decaying particle (mother) decays with some effective strength $y$ into its decay products (daughters), 
an effective decay width might be given as $\G \sim y^2m /(16 \pi)$, where $m$ denotes the mass of the mother.
For the following it is assumed that the branching ratio of this decay is close to one. Actually, we find in Sec.~\ref{sec:brconstraints} that  in all cases we are aware of the branching ratio into dark components is constrained to be very close to one at times later than $t_\text{BBN}\sim 0.1 \seconds$.
Such branching ratios are common in dark matter models, because usually some symmetry is invoked to stabilise the dark matter candidate. 
In the on-shell tree approximation some heavier particle sharing the symmetry then has to decay into the dark matter candidate.
Such branching ratios may also --or in addition-- be enforced by the mass spectrum allowing for only one decay channel kinematically.

A prime example for a dark matter stabilising symmetry is $R$-parity in supersymmetric models, which  also naturally comprise extremely long-lived particles, if combined with gravity.
Think about the gravitino decaying into axino and axion with effective $y^2 \sim \mgrav^2/(12 \mplanck^2)$, where $\mgrav$ denotes the gravitino mass and $\mplanck$ the reduced Planck mass. This decay naturally leads to the emergence of dark radiation way after BBN but before photon decoupling~\cite{Hasenkamp:2011em}.
The mother (gravitino) would decay into a fermionic axino and an axion scalar.
Another example is the decay of the lightest ordinary supersymmetric particle (LOSP) into its superpartner and the gravitino with effective
$y^2 \sim m_\text{losp}^4 / (3\mplanck^2 \mgrav^2)$, where $m_\text{losp}$ denotes the LOSP mass. The scenario reminds of the  sWIMP mechanism, where decays of this kind were considered to produce the observed dark matter.
In the case of a neutralino LOSP the mother were a fermion decaying into the fermionic gravitino and a gauge boson.
In the case of a sneutrino LOSP the mother were a scalar and both daughters (neutrino and gravitino) fermions.
We mention a third example. In higher-dimensional theories the superpartner of a modulus field, i.e., a modulino might decay with effective  $y^2 \sim \l m_{\widetilde \phi}^2 /(3\mplanck^2)$, where $m_{\widetilde \phi}$ denotes the modulino mass and $\l$ a coupling. The modulino might decay into a sneutrino-neutrino or axino-axion pair and so on.
Note that there are various combinations of spins.

\subsection{Basics}
We introduce useful parameters, determine general properties and derive basic equations.
\paragraph{Kinematics}
In the rest frame of the decaying particle the decay products of a two-body decay have in full generality momenta with opposite direction and same absolute value. It is
\begin{equation}
\label{genmomentum}
 |\overrightarrow{p_1}|=|\overrightarrow{p_2}| = \frac{1}{2m} \left( 
(m^2 - (m_1 +m_2)^2)
(m^2 - (m_1 -m_2)^2)
\right)^\frac{1}{2} \, ,
\end{equation}
if $m$ denotes the rest mass of the decaying particle and subscripts 1,2 label the two decay products. We choose subscripts such that $m_1 < m_2$. 
We find it useful to define
\begin{equation}
\label{defdelta}
 \d \equiv \frac{m-m_2}{m_2} = \frac{m}{m_2} -1 >0
\end{equation}
as measure of the mass hierarchy between mother and the heavier daughter or their mass degeneracy for $\d \lesssim 1$.
A negative $\d$ is not possible, because the decay were kinematically forbidden in that case.
The mass of the heavier daughter can be written as
\begin{equation}
\label{m2}
 m_2 = (\d +1)^{-1} m \,.
\end{equation}
If the decay shall produce dark radiation, the energy of the lighter
decay product, $E_1= \sqrt{|\overrightarrow{p_1}|^2 +m_1^2}$, must be
dominated  by its kinetic energy, $E_1 \simeq |\overrightarrow{p_1}|$.
In the limit $m_1 \ll m_2$, which is equivalent to $m_1/m \ll (\d +1)^{-1}$, the general momentum~\eqref{genmomentum} simplifies as
\begin{equation}
\label{p1}
 \lim_{m_1 \ll m_2 } |\overrightarrow{p_1}| = \frac{m}{2} \frac{(\d +1)^2 -1}{(\d +1)^2} 
= \frac{m}{2} \left( 1-\frac{1}{(\d+1)^2} \right)\, .
\end{equation}
For the case $m_1=m_2$ see Sec.~\ref{sec:tddms}.

\paragraph[T2nr]{On $T_2^\text{nr}$}
An initial particle momentum $p_\text{ini} \equiv |\overrightarrow{p}_\text{\!ini}|$ from a decay at 
temperature $\Tdec$ decreases due to the expansion of the Universe. 
Here and in the following, temperatures $T$ refer to the corresponding photon temperature $T_\g$
 at the considered time, e.g., $\Tdec=T_\g(\t)=T(\t)$, where $\t$ is the lifetime of the mother.
  The momentum at temperature $T$ is
\begin{equation}
\label{pofT}
 p(T) = p_\text{ini} \frac{\adec}{a}=p_\text{ini}  \frac{T}{\Tdec}  \left(\frac{\gasts}{\gastsd}\right)^\frac{1}{3} ,
\end{equation}
where the second equality is due to the conservation of comoving entropy.
As usual, $\gasts=\gasts(T)$ is the effective number of degrees of freedom in the entropy density of the Universe
$
s= (2 \pi^2/45) \gasts T^3
$.
The superscript on $\gasts$ indicates here and in the following at which temperature or time $\gasts$ is evaluated, $\gastsd \equiv \gasts (\Tdec)$. The same holds for subscripts on $a$.
We define the temperature $T^\text{nr}$ when a particle with mass $m$ becomes non-relativistic by
\begin{equation}
\label{nrcond}
 p(T^\text{nr}) = m \, .
\end{equation}
For a particle species following the distribution $P$ we consider the mean momentum to determine whether the species is relativistic or non-relativistic. 

In our case $p_\text{ini}$ is given by~\eqref{p1} and the mass of the heavier daughter by~\eqref{m2}.
Therefore, the condition~\eqref{nrcond}
yields
\begin{equation}
\label{T2nr}
 T_2^\text{nr} = \Tdec \frac{2}{\m} \frac{ \d +1}{(\d +1)^2 -1} \left(\frac{\gastsd}{\gastsnr}\right)^\frac{1}{3} 
\end{equation}
with $\gastsnr\equiv\gasts(\Tnr)$.
The correction factor $\mu=c^{-1}\G[c^{-1}]$ takes into account the exponential decay law in an expanding universe, $a\propto t^{1/c}$, compared to the sudden decay approximation. It is derived in Appendix~\ref{appendix:expdecay}.
It is $\m=\mu(P) \simeq 0.886$ if the decay occurs during radiation domination and $\mu\simeq 0.902$ if the decay occurs during matter domination. 
Throughout this work we will often argue under the assumption of a
sudden decay, because this simplifies the discussion and reveals key
points. One example for this is considering some notion ``at decay''.
We will take into account corrections due to the exponential decay law in the final equations by correction factors, which represent good approximations for times $t \gtrsim 3\t$ or $ \gtrsim 4\t$, cf.~Appendix~\ref{appendix:expdecay}. Often times of interest are indeed much later than the time of decay.

\paragraph{Energy densities}
The ``non-dark'' radiation energy density of the Universe, i.e.,
the energy density of the thermal bath in the Universe, is given by
\begin{equation}
\label{rhorad1}
 \rho_\text{rad} = \frac{\pi^2}{30} g_\ast T^4 \,,
\end{equation}
where $\gast=\gast (T)$ denotes the effective number of relativistic degrees of freedom in the bath.
Bounds on the total radiation energy density $\r_\text{rad}^\text{tot}$
 exist from processes around and after $e^+e^-$-annihilation, so for cosmic temperatures around and smaller than the $e^+e^-$-annihilation temperature $T_{e^+e^-}\sim m_e \simeq 0.5 \MeV$.  They are usually given in terms
 of the effective number of neutrino species $\Neff$ defined by
 \begin{equation}
  \r_\text{rad}^\text{tot} = \left( 1 + \Neff \frac{7}{8} \left(\frac{T_\n}{T}\right)^4 \right) \r_\g \, ,
  \label{rhorad}
 \end{equation}
where the radiation energy density is given as a sum of the
energy density in photons $\r_\g=(\pi^2/15) T^4$, the energy density in SM neutrinos with $\Neff^\text{SM}=3.046$~\cite{Mangano:2005cc}
and $T_\n/T = (4/11)^{1/3}$ and
any departure from the standard scenario parametrised as a summand in $\Neff = \Neff^\text{SM} + \D\Neff$.
The small deviation of $\Neff^\text{SM}$ from $3$ is due to incomplete neutrino decoupling at $e^+e^-$-annihilation. 
We denote temperatures before neutrinos become non-relativistic at
$T_\nu^\text{nr}$ and lower than $T_{e^+e^-}$ by $\Tlow$.
Comparing~\eqref{rhorad1} and~\eqref{rhorad} we see that
$\gast(\Tlow)\simeq 3.384$.\footnote{
The differences in $\gast$ and $\gasts$ to the often used values in the
literature are due to $\Neff^\text{SM} \neq 3$.  Note that by definition
we do not consider the daughter particles in the determination of
$\gast$. Furthermore, their entropy is negligible, so they do not
change $\gasts$.
}
In this temperature range the energy density in dark radiation $\rhodr$ can be written as
\begin{equation}
 \rhodr = \D\Neff \times \r_{1\nu} = \D\Neff \times \frac{7}{8} \left(\frac{T_\nu}{T_\g}\right)^4 \r_\g \, ,
\end{equation}
 if $\r_{1\nu} $ denotes the energy density of one SM neutrino species with thermal spectrum.
It follows that
\begin{equation}
\label{rhodrlow}
 \left.\frac{\rhodr}{\rho_\text{rad}^\text{SM}}\right|_\text{low} = 0.1342 \times \D\Neff 
\end{equation}
for $T_\nu^\text{nr} < T <T_{e^+e^-}$.
We see that even for the 5-$\s$ limit, $\D\Neff^\text{max}=5.265$, of the combined analysis in~\cite{Dunkley:2010ge},
there would be less dark than SM radiation.
Here and in the following a vertical line with subscripts like $\left.\right|_\text{low}$ indicates at what time (or temperature) the corresponding term is evaluated.
Towards higher temperatures the bath energy density $\r_\text{rad} \propto g_\ast T^4$, while the one in dark radiation scales as $\rhodr \propto g_{\ast s}^{4/3} T^4$, if dark radiation is decoupled from the bath.
Thus, at the time of
the decay producing the dark radiation,
\begin{equation}
\left.\frac{\rhodr}{\r_\text{rad}}\right|_\text{dec} = 
\frac{\rhodr (T_\text{dec})}{\rhodr (\Tlow)}
\frac{\rhorad^\text{SM}(\Tlow)}{\rhorad (T_\text{dec})}
\left.\frac{\rhodr}{\rho_\text{rad}^\text{SM}}\right|_\text{low}
 = \left(\frac{\gastsd}{\gasts^\text{low}}\right)^\frac{4}{3}
\left(\frac{\gast^\text{low}}{\gastd}\right)
\left.\frac{\rhodr}{\rho_\text{rad}^\text{SM}}\right|_\text{low} ,
\end{equation}
and inserting $g_\ast(\Tlow)$ and
$g_{\ast s}(\Tlow) = \gastst = 2 (1+ \Neff^\text{SM} 28/88) \simeq 3.938$
finally yields\footnotemark[\thefootnote]
\begin{equation} 
\label{rhodratdec}
\left.\frac{\rhodr}{\r_\text{rad}}\right|_\text{dec}
 = 0.5440  \frac{(\gastsd)^{4/3}}{\gastd} 
\left.\frac{\rhodr}{\rho_\text{rad}^\text{SM}}\right|_\text{low} \, ,
\end{equation}
which is valid at any decay temperature.
Due to the different scaling behaviour the dark radiation component could have dominated the Universe at decay, but only for decays with $\gast=\gasts \gg \gastt$ and for extreme values of $\D\Neff$.
At intermediate temperatures $T$ we use the different scaling of dark
and SM radiation to derive
\begin{equation}
\label{rhodr}
 \rhodr(T) = 0.0730 \D\Neff \frac{\gasts^{4/3}}{\gast} \rhorad(T) \, .
\end{equation}
This is a useful parametrisation of the dark radiation energy density.

The desired amount of dark radiation determines the energy density of the decaying particle $\r= n\, m$  at its decay.
In a two-body decay with branching ratio one the number densities of the decay products are fixed to be equal to the number density of the decaying particle, $n=n_1=n_2$.
As radiation the energy of a particle can be approximated by its momentum, $E\simeq p$, and the heavier daughter may act or may not act as dark radiation at the times of observation. Then the energy density of dark radiation at decay reads 
$ 
\rhodr|_\text{dec} \simeq \gdrobs n E_1 \simeq \gdrobs n p_1 
$,
where $p_1$ is determined from the kinematics~\eqref{p1} and $\gdrobs$
counts the number of dark radiation components  at the time of
observation. It is $\gdrobs=1$, if the heavier daughter particle became
non-relativistic before the time probed by observations, and $\gdrobs=2$
otherwise.
We define a conversion factor $f$ by 
\begin{equation}
\label{deff}
\rhodr = f \times \r 
\end{equation}
such that
\begin{equation}
\label{f}
 f= \mu \frac{\gdrobs}{2} \frac{(\d +1)^2-1}{(\d +1)^2} \, .
\end{equation}
Altogether, we obtain
\begin{equation}
\label{rhoatdec}
 \left.\frac{\r}{\rhorad}\right|_\text{dec} = f^{-1} \left.\frac{\rhodr}{\rhorad}\right|_\text{dec}
= \frac{0.146}{\m} \frac{\D\Neff}{\gdrobs} \frac{(\gastsd)^{4/3}}{\gastd} \frac{(\d+1)^2}{(\d+1)^2-1} \, .
\end{equation}
We see that the decaying particle could be required to dominate the energy density of the Universe, $\r >\rhorad$,
 for a short time prior to its decay to explain extreme values of $\D\Neff$. However, this is not to be expected and, especially at late times, improbable
since a large $\gast \gg \gast^\text{SM}$ would be necessary.
From~\eqref{rhoatdec} follows today's energy density of the decaying
particle, if it had not decayed, in units of today's critical energy density $\rhocrit$ as
\begin{equation}
\label{omegamother}
 \O h^2 = \frac{\rho|_\text{dec} h^2}{\rhocrit} \left(\frac{T_0}{\Tdec}\right)^3 \frac{\gastst}{\gastsd} \, ,  
\end{equation}
where the dilution is considered that would have happened till today.
In this work we will heavily use the fact that the energy density of the
decaying particle is fixed by the amount of dark radiation and thus by
observations, independent of an underlying particle physics model.
The other way around we can single out $\D\Neff$ in~\eqref{rhoatdec} resulting in
\begin{equation}
 \D\Neff = 9.13 \mu \frac{mY}{\Tdec}
\frac{(\d+1)^2-1}{(\d+1)^2}
\frac{\gdrobs}{(\gastsd)^{1/3}}   \, ,
\end{equation}
where we introduced the particle yield $Y \equiv n/s$. 

We note in passing that the decaying particle is allowed to dominate the
Universe prior to its decay, if it decays before BBN,
$\t \ll t_\text{BBN}$.  In this case $\D\Neff$ is set by the relative
branching into dark radiation and radiation formed by SM particles.
This relative branching is given by the ratio
$\rho_\text{dr}/\rho_\text{rad} = 0.073 \, \D\Neff \, \gast^{1/3}$ found
in~\eqref{rhodr}.

\subsection[delta-tau plane]{$\delta$-$\tau$ plane}
\label{sec:delta-tau-plane}
\begin{figure}
 \centering
   \includegraphics[width=\textwidth]{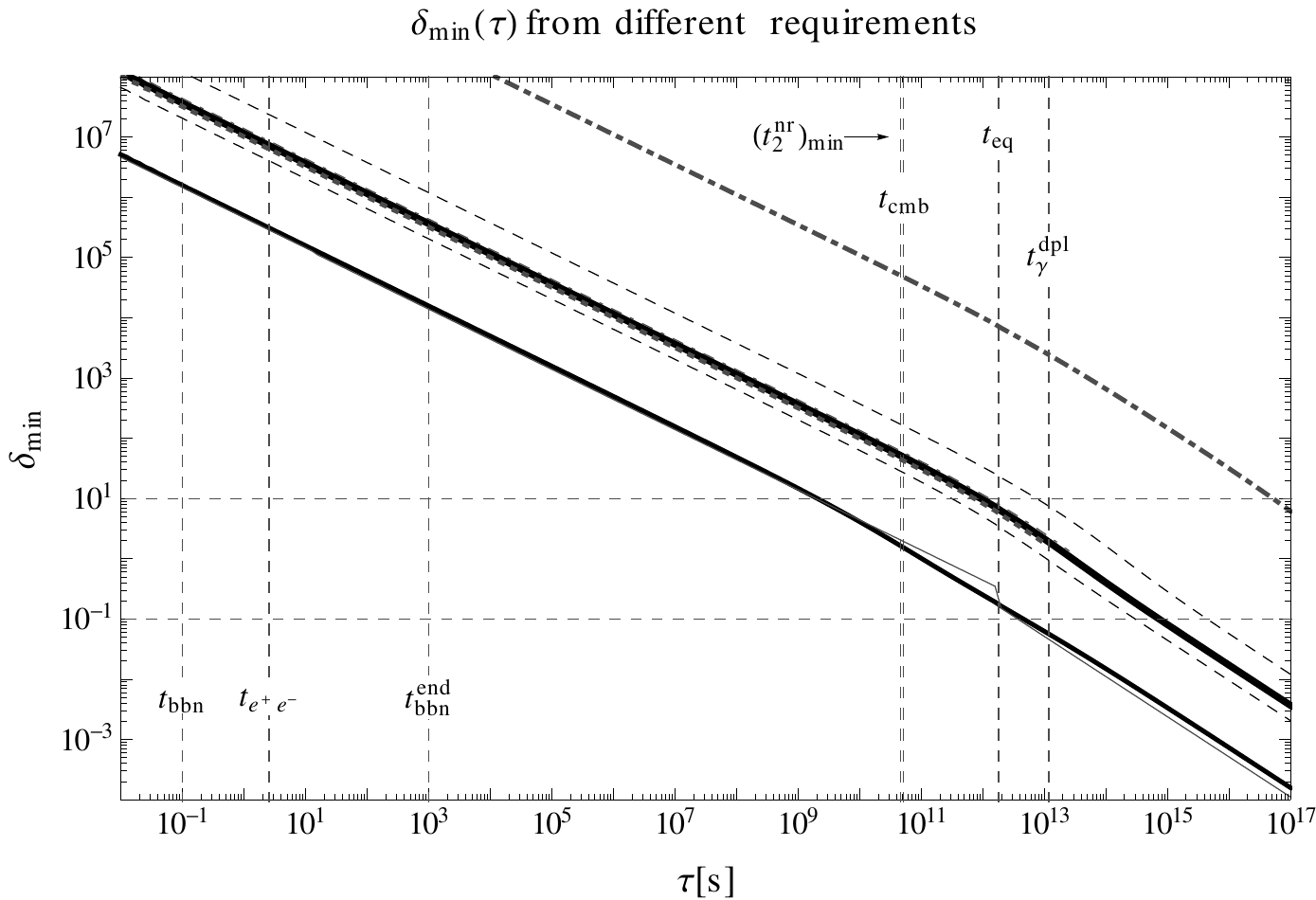}
\caption[$\d_\text{min}$-$\t$ plane]{$\d_\text{min}$-$\t$-plane exploiting~\eqref{implicitlowboundondelta}.
Values above the corresponding line are considered to be allowed.
 The thick solid curve corresponds to the hot dark matter constraint, cf.~Sec.~\ref{sec:hdm}, with $\D\Neff =1$. Thin dashed curves below ($\D\Neff=0.52$) and above ($\D\Neff=5.265$) show the dependence of this bound on the produced amount of dark radiation.
The thin solid curve corresponds to the non-domination constraint.
The analytic approximations~\eqref{lowboundondelta} and~\eqref{lowboundondeltaafter} are overplotted as very thin grey curve with a jump at $\teq$.
 The dotted curve (at the lower edge of the thick solid one) gives $\d$ such that the heavier daughter becomes non-relativistic at photon decoupling $t_\g^\text{dpl}$ and the dash-dotted one such that this happens today. These three curves are for $\D\Neff=1$.
At the upper edge of the thick solid curve mean values of~\cite{Benson:2011ut} originate from the decay for massless neutrinos.
Various important and suggestive times are highlighted by vertical dashed lines: onset $t_\text{bbn}$ and end of BBN $t_\text{bbn}^\text{end}$, the earliest possible time for the heavier daughter to become non-relativistic $\tnrmin$ from~\eqref{tnrmin}, re-entry of first observable modes in the CMB $\taumax$ and matter-radiation equality $\teq$.
For decays during BBN the increase in $\Neff$ determined from BBN is
smaller than the corresponding increase measured in the CMB\@. The relative difference depends on the time of decay as quantified in~\cite{Menestrina:2011mz}.
In all figures we take into account the evolution of $\gast$ and
$\gasts$. Nevertheless, the curves are smooth around $e^+e^-$
annihilation at $t_{e^+e^-}$, which shows that dependencies have cancelled.
Within the horizontal dashed lines the mass hierarchy or degeneracy is within an order of magnitude.
}
\label{fig:delta-tau-dr}
\end{figure}
The mass hierarchy between decaying particle and decay products is
constrained by several cosmological considerations.
First of all, at no time there is an upper bound on $\d$, because the energy density of the decaying particle can, in principle, be adjusted such that all its energy transferred to radiation at its decay accounts for the desired increase in $\D\Neff$.
This might be different in a concrete particle physics model where $\d$
and/or $\r$ are given.
More importantly, there are lower bounds on $\d$ from cosmology.
The underlying considerations are qualitatively different depending on
whether the decay occurs before or after matter-radiation equality at $\teq$.
In any case they depend on the time of decay $\t$, while they are independent of the underlying particle physics model.
Our assumptions on the time and temperature of equality are outlined at the beginning of Sec.~\ref{sec:structureformation}.

\textbf{Before $\teq$} there is a lower bound on $\delta$ from the requirement that no daughter
particle may come to dominate the Universe before $\teq$. This non-dominance requirement can be expressed as
\begin{equation}
\label{nondom}
 \O_2 h^2 \leq \bmax \omegadm  
\end{equation}
with $\bmax = 1$.
When the mother decays (not too close before $\teq$) also the heavier daughter particle must be emitted as radiation. 
Otherwise, it would dominate already shortly after being emitted,
since $\r$ has to make up a sizeable fraction of the total energy density,
 see~\eqref{rhoatdec} and Fig.~\ref{fig:energydensities}.
 Thus its energy density is as large as the energy density of the lighter daughter, $\r_2|_\text{dec} =\r_1|_\text{dec} = (\gdrobs)^{-1} \rhodr(\Tdec)$. 
After emission its energy density scales as radiation $\propto a^{-4}$ till it becomes non-relativistic at $T_2^\text{nr}$. 
If $T_2^\text{nr}< \Teq$, $\rho_2$ surely never dominates.
From becoming non-relativistic on, it scales as matter $\propto a^{-3}$. Thus
\begin{equation}
 \O_2 = \frac{\rhodr(\Tdec)}{\rhocrit} (\gdrobs)^{-1} 
\frac{T_2^\text{nr} T_0^3}{\Tdec^4}
 \left(\frac{\gastst}{\gastsd}\right)^\frac{4}{3} \, ,
\end{equation}
where we anticipated $\gastsnr=\gastst$.
Inserting $\rhodr$ from~\eqref{rhodr}, $\rhorad$ from~\eqref{rhorad1} as
well as known numerical values \cite{Beringer:1900zz}, the inequality~\eqref{nondom} becomes an upper bound on $T_2^\text{nr}$
\begin{equation}
\label{T2nrbound}
T_2^\text{nr} \leq 7.161 \, \Teq \D\Neff^{-1} \bmax \left(\frac{\omegadm}{0.1286} \right) \gdrobs \,,
\end{equation}
where we can see that for the bound only $\gdrobs=1$ is sensible.
Since $T_2^\text{nr} \ll T_{e^+e^-}$, it is justified to set $\gastnr=\gastt$.
We find the corresponding cosmic time as
\begin{equation}
\label{tnrmin}
\tnrmin = 0.0238 \, \teq \, b_\text{max}^{-2} \D\Neff^2 \left(\frac{0.1286}{\omegadm}\right)^2 =
4.33 \times 10^{10} \!\seconds \; b_\text{max}^{-2} \, \D\Neff^2 \left(\frac{0.1286}{\omegadm}\right)^2 .
\end{equation}
For times later than $\tnrmin$ the assumption of relativistic emission is  no longer necessarily fulfilled.
We note that $t_2^\text{nr}$ corresponding to $T_2^\text{nr}$ from~\eqref{T2nrbound}
is likely later than the time when the first observable modes of the CMB enter the horizon, which sets $\taumax \simeq 5.2 \times 10^{10} \seconds $~\cite{Fischler:2010xz}.
Indeed, this is for sure taking into account constraints from structure formation requiring $b_\text{max}< 1$, cf.~Sec.~\ref{sec:structureformation}.
The heavier daughter is restricted to become non-relativistic during CMB times or later, which might leave observable consequences due to the corresponding change in the expansion rate.
Likewise, for non-relativistic emission in the intermediate regime, $\tnrmin < \t \lesssim \teq$, we expect observable consequences in the CMB. 

Inserting $T_2^\text{nr}$ from~\eqref{T2nr} into~\eqref{T2nrbound} we obtain an implicit lower bound on $\d$
\begin{equation}
\label{implicitlowboundondelta}
\frac{(\d+1)^2-1}{\d+1} > \frac{0.2793}{\m} \frac{\Tdec}{\Teq} \D\Neff 
\left(\frac{0.1286}{\omegadm}\right)
\left(\frac{\gastsd}{\gastst}\right)^\frac{1}{3} .
\end{equation}
If $\d \gg 1$, the l.h.s.\ of~\eqref{implicitlowboundondelta} reduces as
\begin{equation}
\label{largedeltalim}
 \lim_{\d \to \infty} \frac{1}{\d}\frac{(\d+1)^2-1}{\d+1} = 1 \, .
\end{equation}
Then~\eqref{implicitlowboundondelta} becomes practically
\begin{equation}
\label{lowboundondelta}
\d > \frac{0.2793}{\m} \left(\frac{\teq}{\t}\right)^\frac{1}{2} \D\Neff 
\left(\frac{0.1286}{\omegadm}\right)
\left(\frac{\gastt}{\gastd}\right)^\frac{1}{4}
\left(\frac{\gastsd}{\gastst}\right)^\frac{1}{3}
,
\end{equation}
where we used the time-temperature relation in a radiation-dominated universe to replace
$\Tdec/\Teq = (\gasteq/\gastd)^{1/4} (\teq/\t)^{1/2}$
and $\gasteq= \gastt$.

\textbf{After $\teq$} a relativistically emitted, non-dominating particle becomes even more subdominant as the Universe expands.
However, there is a lower bound on $\d$, if we require some significant
$\D\Neff > 0$, because the maximally allowed energy density for the
heavier daughter particle is the dark matter energy density
as in~\eqref{nondom} with the crucial difference that it is emitted
non-relativistically if it saturates the bound.
If $\bmax \simeq 1$, dark matter (the mother) would decay
and convert a small amount of its energy into radiation.
This is qualitatively different for $\t < \tnrmin$.

The energy density of a non-relativistic species can be approximated as
$
 \r_2= n_2 E_2 \simeq n_2 m_2
$.
Again exploiting $n =n_2 = n_1$ and replacing $m_2$ by~\eqref{m2} we find
\begin{equation}
 \r_2 = \frac{n\, m}{\d+1} \Rightarrow  \r_2|_\text{d} = (\d+1)^{-1} f^{-1} \rhodr(\Tdec) \, ,
\end{equation}
where we used that $\r=n m$ and the definition of the conversion
factor~\eqref{deff}. Taking into account the expansion till today the energy density of the heavier daughter in units of today's critical energy density is given by
\begin{equation}
\O_2 h^2 = \frac{2}{\m} \frac{\d +1}{(\d +1)^2 -1} \frac{\rhodr (\Tdec) h^2}{\rhocrit} \left(\frac{T_0}{\Tdec}\right)^3 .
\end{equation}
From the requirement~\eqref{nondom} we obtain the very same implicit lower bound~\eqref{implicitlowboundondelta} on $\delta$ for decays after $\tnrmin$, where only $\gastsd/\gastst$ has to be replaced by one.
If $\d \rightarrow 0$, the l.h.s.\ of~\eqref{implicitlowboundondelta} reduces as
\begin{equation}
\label{smalldeltalim}
 \lim_{\d \to 0} \frac{1}{2\d}\frac{(\d+1)^2-1}{\d+1} = 1 \, .
\end{equation}
Then~\eqref{implicitlowboundondelta} becomes practically
\begin{equation}
\label{lowboundondeltaafter}
\d > \frac{0.2793}{2\m} \left(\frac{\teq}{\t}\right)^\frac{2}{3} \D\Neff 
\left(\frac{0.1286}{\omegadm}\right) ,
\end{equation}
where we used the time-temperature relation in a matter-dominated universe with constant $\gast$ to replace
$
 \Tdec/\Teq = (\teq/\t)^\frac{2}{3} 
$.
Whenever necessary we assume a sudden transition from radiation to matter domination in analytic calculations.

The lower bound on $\d$ from~\eqref{implicitlowboundondelta} with $\D\Neff=1$ is depicted in Fig.~\ref{fig:delta-tau-dr} as thin solid curve.
The analytic approximations~\eqref{lowboundondelta} and~\eqref{lowboundondeltaafter} are overplotted as very thin grey curve with its largest deviation at $\teq$ and a tiny underestimation at very late times.
A decay before $t_\text{bbn}$ increases $\D\Neff$ before BBN\@.
We found that the decay products may not become non-relativistic before
a time $\tnrmin \sim \taumax$.
For decays during BBN the increase in $\Neff$ determined from BBN is smaller than the increase measured in the CMB depending on the time of decay~\cite{Menestrina:2011mz}. 
Decays after $t_\text{bbn}^\text{end}$ add radiation during CMB times. It is usually assumed that the decay products are still relativistic today or at least till photon decoupling at $t_\g^\text{dpl}$, while this need not be the case, in particular for the heavier daughter.
The figure ranges beyond $t_\g^\text{dpl}$ to times as late as $10^{17} \seconds$. This is for completeness. For such late decays the meaning of the curve may be far from clear in this and following graphs due to the currently limited understanding.
Nevertheless, first pieces of information are provided that might motivate focused studies of such cosmologies.

Let us conclude the discussion with a number of comments.
We could have started from the general requirement~\eqref{nondom} for any time and would have obtained the same results. We would like to repeat that close to the boundary the cosmology is very different for a decay occurring sufficiently before or after $\tnrmin$.
After $\tnrmin$ the energy density of the decaying particle is of the order of the energy density of the heavier daughter. In some sense dark matter decays into today's dark matter particle and dark radiation.
Before $\tnrmin$ the energy density of the decaying particle has to be much larger than the dark matter energy density to produce significant dark radiation.
One could expect a lower bound on $\d$ from requiring some significant $\D\Neff$ to arise also before $\tnrmin$. 
However, this is not the case. The energy density of the decaying particle could in principle be arbitrarily large, such that in its decay a tiny amount of its energy converted into dark radiation suffices. This is because dark radiation does not thermalise and thus does not dilute pre-existing abundances. The additional entropy is negligible as the momenta of the decay products are distributed on a small shell in phase space, cf.~Appendix~\ref{appendix:expdecay}. An arbitrarily small $\d$ seems allowed.
However, the non-dominance requirement of~\eqref{nondom} were violated. For $\d$s fulfilling~\eqref{implicitlowboundondelta} the upper bound on the energy density of the decaying particle is set by the possible overproduction of dark radiation, 
i.e., a too large increase in $\D\Neff$ inconsistent with observations.
In the scenario under consideration this upper bound were set via~\eqref{rhoatdec}.
The lower bound on $\d$ requiring some minimal $\D\Neff$ is obtained by inserting this value into~\eqref{implicitlowboundondelta}.
For times around $\teq$ the situation is more involved than outlined here, because quite a large fraction of the dominating matter component would have to decay into radiation. However, we expect such a situation to leave pronounced signatures in the CMB.

\subsection{Structure formation}
\label{sec:structureformation}
The transition from radiation to matter domination is one of the most important events in structure formation. The red-shift of matter-radiation equality $\zeq$ is related to the dark radiation and matter density $\O_\text{m}$ by (cp.~(53) of~\cite{Komatsu:2010fb})
\begin{equation}
\label{zeq}
 1+\zeq = 3201
\left( \frac{\D\Neff}{7.44} +1 \right)^{-1}
\frac{\O_\text{m} h^2}{0.1333} \, .
\end{equation}
Naively, one could conclude that $\D\Neff > 0$ thus implies smaller $\zeq$, i.e., a later transition. However, this is not observed.
What can instead be seen in the CMB power spectrum is a suppression at higher multipoles (corresponding to smaller scales) compared to the expectation from the standard scenario. Constraints on $\Neff$ depend on the cosmological model, which can be extended in other ways to suppress small-scale power. For an increase in $\Neff$ the suppression has been identified to be due to increased Silk damping~\cite{Hou:2011ec}.
 Indeed,~\eqref{zeq} shows the degeneracy of $\Neff$ and $\O_\text{m} h^2$, in particular, for WMAP using the first and third acoustic peak to determine $\zeq$. This degeneracy is broken by including measurements on smaller scales. The combined data then allows to measure $\Neff$ in addition to $\zeq$, see references in the introduction. The studies do not find that $\zeq$ varies with $\D\Neff$. Even though they also do not exclude a smaller $\zeq$, we adopt the PDG mean value $\zeq =3200 \pm 130$ \cite{Beringer:1900zz}, which is consistent with the other studies.
The temperature at equality is thus $\Teq = (1+\zeq) T_0 \approx 0.752
\eV$. We fix the time of matter-radiation equality to $\teq \approx 1.81
\times 10^{12} \seconds$ from the relation between time and red-shift in
a universe filled with radiation and matter\footnote{
In all numerical calculations, we employ this relation rather than the
approximate relations valid in the limit of complete radiation or matter
domination.}
using PDG mean values for the time and red-shift of decoupling~\cite{Beringer:1900zz}. 

For a fixed $\zeq$ the matter and radiation density are no longer independent.
In Fig.~\ref{fig:energydensities} this can be seen comparing the energy density of the Universe including dark radiation (thick, solid) with standard cosmology (grey, dashed).
The baryon density is regarded as robustly measured, $\O_\text{b} h^2 =0.0226(6)$ with $1\s$ uncertainty in the last digit~\cite{Beringer:1900zz}. Therefore, the uncertainty in $\D\Neff$ turns into an uncertainty in the dark matter density as  $\O_\text{m} = \O_\text{b} + \O_\text{dm}$.
Following~\eqref{zeq} dark matter and dark radiation energy densities are linked as
\begin{equation}
 \omegadm = 0.1107 + 0.0179 \D\Neff \, .
\end{equation}
Omitting this dependence in, for example, Sec.~\ref{sec:delta-tau-plane}
can lead to deviations of up to some ten percent.
Thus it is important to take this dependence into account. This is especially true when considering more complete cosmologies including, for example, an origin of dark matter as in Sec.~\ref{sec:tddms}. Particle physics parameters can be significantly affected. 

Increasing the dark matter energy density, while keeping the baryon density constant, decreases the baryonic matter fraction $\O_\text{b}/\O_\text{m}$. 
This decreases the pressure support on matter prior to photon decoupling and, therefore, boosts the growth of structures below the sound horizon at photon decoupling of about $150 \Mpc$~\cite{Eisenstein:1997ik}.
A finite neutrino mass has the opposite effect, because neutrinos then
count towards $\O_\text{m}$, while they do not take part in structure formation below their free-streaming scale.
Consequently, a best-fit might be found having both, additional radiation and a --small, ``compensating''-- dark matter fraction with very large free-streaming scale, see~\cite{Hamann:2011ge,Joudaki:2012uk} considering sterile neutrinos.
In this line of thought it is very interesting that $\D\Neff>0$ together with a finite neutrino mass scale seems to be favoured by measurements of the abundance of galaxy clusters corresponding to $\sim 10 \Mpc$ scales~\cite{Benson:2011ut}. 
We note that all these considerations assume a decay before the affected
epoch, for example, before $\teq$ to increase the amount of radiation at $\teq$ and so on. It is neither clear how an increase in $\Neff$ after $t_\g^\text{dpl}$ might be observed nor what is the effect of the corresponding amount of matter being transformed into radiation. Also in this sense some figures range beyond $t_\g^\text{dpl}$ for completeness only.

In the following we shall point out the constraints and opportunities arising from the heavier daughter in structure formation. 
It may constitute just half of the dark radiation, but it does not need to act as radiation at all. The opposite extreme case would be that the heavier daughter forms the observed dark matter. In between these two cases its free-streaming could mimic the finite neutrino mass scale as deduced from cosmological observations.

\subsubsection{Minimal free-streaming scale}
\begin{figure}[ht]
 \centering
   \includegraphics[width=0.8 \textwidth]{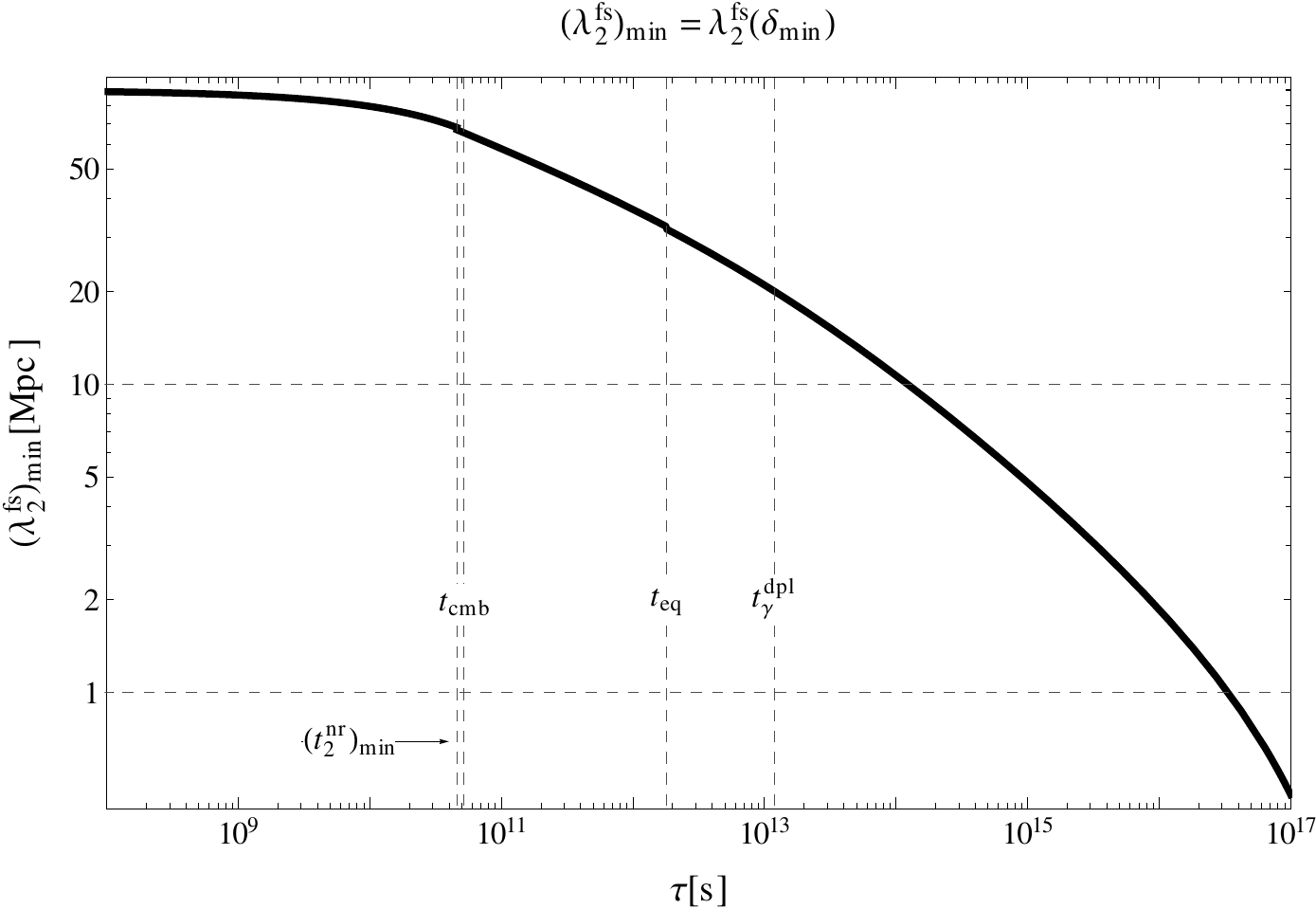}
\caption{Minimal free-streaming scale of the heavier daughter $(\l_2^\text{fs})_\text{min}$ depending on the time of decay for $\D\Neff=1$.
Times are highlighted as in Fig.~\ref{fig:delta-tau-dr}.
Horizontal dashed lines indicate the galaxy cluster scale $\l_\text{gc} \sim 10 \Mpc$ and $1 \Mpc$ as suggestive free-streaming scale of warm dark matter, respectively.
}
\label{fig:lfsmin1}
\end{figure}
As collisionless particle the heavier daughter can stream out of
overdense regions and into underdense regions, smoothing out
inhomogeneities.  In order to take this effect into account precisely,
one has to integrate Boltzmann equations. In the following we shall estimate the scale of collisionless damping analytically.
The decisive quantity is the free-streaming scale
\begin{equation}
\label{lfs1}
  \l_2^\text{fs} \equiv \int_\t^{t_0}  \frac{v_2(a)}{a} dt
\end{equation}
of the heavier daughter, where $v(a)$ denotes velocities that depend on
the scale factor and where for convenience we have chosen $a_0=1$ for
today's scale factor.
The situation is qualitatively different depending on whether the decay occurs before or after $\tnrmin$.

\textbf{Before $\tnrmin$} both decay products are emitted relativistically.
If a particle becomes non-relativistic after $\teq$, its free-streaming scale is much larger than the size of a protogalaxy. It is said to act as hot dark matter, see Sec.~\ref{sec:hdm}.
We will be interested in the case used in Sec.~\ref{sec:delta-tau-plane} to determine $\d_\text{min}$ where the heavier daughter becomes non-relativistic before $\teq$. 
We approximate the velocity of a particle emitted with relativistic momentum by 
\begin{equation}
 v(a)\approx \begin{cases}
        1 & \text{, if } a<a_\text{nr} \\
\frac{a_\text{nr}}{a} & \text{, if } a \geq a_\text{nr} \, .
       \end{cases} 
\end{equation}
After the emission with $v\simeq 1$ the velocity decreases with the expansion of the Universe and becomes smaller than one when the particle becomes non-relativistic at $a_\text{nr}$.
A finite, non-relativistic velocity of the decaying particle is safely negligible.
Exploiting the additivity of integration on intervals we thus find
\begin{equation}
  \l_\text{fs} \approx \int_\t^{t_\text{nr}} \frac{1}{a} dt
+ \int_{t_\text{nr}}^{\teq} \frac{a_\text{nr}}{a^2} dt 
 + \int_\teq^{t_0} \frac{a_\text{nr}}{a^2} dt \, .
\end{equation}
The first integral corresponds to relativistic free-streaming, the second one to non-relativistic free-streaming before $\teq$ and the third to non-relativistic free-streaming during the matter-dominated era. We can safely neglect the current vacuum-dominated phase. 

In order to perform the integration analytically
we take $\gast$ and $\gasts$ to be constant for times as late as
$t_\text{nr}$. As the variation of $\gasts^{1/3}$ and $\gast^{1/4}$ with
time is weak, we treat them like constant factors in the first integral,
too. We checked numerically that this induces negligible errors. 
With the additional approximation of a sudden transition between radiation and matter domination at $\teq$
the integrations become straightforward exploiting the common $a(t)$ relations. We obtain
\begin{equation}
\label{lfsintegrated}
 \l_\text{fs} \approx 2 \frac{\t^\frac{1}{2} t_\text{nr}^\frac{1}{2}}{\adec} 
\left(\frac{\gastsnr}{\gastsd}\right)^\frac{1}{3}
\left(\frac{\gastd}{\gastnr}\right)^\frac{1}{4}
\left(1 - \left(\frac{\t}{t_\text{nr}}\right)^\frac{1}{2} \right)
+ \frac{t_\text{nr}}{a_\text{nr}} \left( 3 + \ln \frac{\teq}{t_\text{nr}}\right) 
\end{equation}
with $\gastsnr = \gasts(t_\text{nr})$ and $\gastnr= \gast(t_\text{nr})$.
The first term corresponds to relativistic free-streaming, the first summand in the brackets of the second term to non-relativistic free-streaming during the matter-dominated era and the logarithm to non-relativistic free-streaming before $\teq$.
Accepting an error at the percent level we have taken $(\teq/t_0)^{1/3} \rightarrow 0$ in the last integral. 
In~\eqref{lfsintegrated} we can see that for the often discussed case $\t\ll t_\text{nr} \ll \teq$ relativistic free-streaming is negligible and non-relativistic free-streaming before $\teq$ dominates over free-streaming after $\teq$.
In this case $\l_\text{fs}$ is roughly set by $t_\text{nr}$. 
In order to investigate our case we set $t_\text{nr}= t_2^\text{nr}= t(\Tnr)$ in~\eqref{lfsintegrated}. 
The scale factors at corresponding times are given by 
$
\adec =
(T_0/\Tdec) (\gastst/\gastsd)^{1/3}
$
and
$
a_\text{nr} =
(T_0/\Tnr) (\gastst/\gastsnr)^{1/3}
$,
respectively. 
Since $T_2^\text{nr}$ is given by~\eqref{T2nr} we can express the
free-streaming scale of the heavier daughter as a function of $\d$ and the time of decay $\t$. We find
\begin{eqnarray}
\label{lfsdeltatau}
 \l_2^\text{fs}(\d,\t) &\approx& 0.09 \Mpc 
\left(\frac{\t}{10^7 \seconds}\right)^\frac{1}{2}
\frac{(\d+1)^2 -1}{ \d+1}
\frac{(\gastd)^{1/4}}{(\gastnr)^{1/2}}
\left(\frac{(\gastsnr)^2}{\gastsd \gastst}\right)^\frac{1}{3}  \nonumber \\
& \times & \Bigg\{
5
- \frac{4}{\mu} \frac{ \d+1}{(\d+1)^2 -1}   
\left(\frac{\gastnr}{\gastd}\right)^\frac{1}{4}
\left(\frac{\gastsd}{\gastsnr}\right)^\frac{1}{3}  \nonumber \\
 &+& \ln \left\lbrack 
\frac{\teq}{\t} \frac{4}{\m^2} 
\left(\frac{ \d+1}{(\d+1)^2 -1}\right)^2
\left(\frac{\gastnr}{\gastd}\right)^\frac{1}{2}
\left(\frac{\gastsd}{\gastsnr}\right)^\frac{2}{3}
\right\rbrack 
\Bigg\} \, .
\end{eqnarray}
This analytic approximation applies to any case with $\t < t_\text{nr} < \teq \ll t_0$.
Of course, as $\t \rightarrow t_\text{nr}$ there is no relativistic free-streaming.

For sufficiently early decays
the minimal free-streaming scale of the heavier daughter
$(\l_2^\text{fs})_\text{min}$ becomes actually independent of the time of
decay. It is instead given by the earliest possible time for the heavier
daughter to become non-relativistic,~\eqref{tnrmin}.  It is
reached for the minimal $\d$ given by~\eqref{implicitlowboundondelta},
because for the minimally required mass hierarchy the heavier daughter
is emitted with its minimal initial momentum.
We find
\begin{eqnarray}
\label{lfsmin}
 (\l_2^\text{fs})_\text{min} (\D\Neff, \t)  &\approx&  10 \Mpc \;
\D\Neff \left(\frac{0.1286}{\omegadm}\right) \nonumber \\
 &\times&\Bigg( 5 - 2 \left(\frac{\t}{(t_2^\text{nr})_\text{min}}\right)^\frac{1}{2} 
 + \ln \left[ 51 \left(\D\Neff^{-1} \frac{\omegadm}{0.1286}\right)^2 \right] \Bigg) \, .
\end{eqnarray}
Consequently it depends on the amount of dark radiation only, except for the relativistic free-streaming scale.
It is independent of $\gast$ and $\gasts$ at decay.
Taking $\D\Neff=1$ and the limit $\t \ll t_\text{nr}$ we obtain
$
 (\l_2^\text{fs})_\text{min} \approx 91 \Mpc
$
as shown in Fig.~\ref{fig:lfsmin1}, which should be compared to the galaxy cluster scale $\l_\text{gc}\sim 10 \Mpc$. 
Thus, the heavier daughter from a decay before $\tnrmin$ is in any case way too warm to form the observed dark matter, see below.

\textbf{After $\teq$} to act as matter the heavier daughter should be emitted non-relativistically. We approximate its velocity by
\begin{equation}
 v_2(a) \approx \frac{p_\text{ini}}{m_2} \frac{\adec}{a} \, ,
\end{equation}
where $p_\text{ini}$ is given by~\eqref{p1}. With $a\propto t^{2/3}$ in a matter-dominated universe the scale factor at decay $\adec=(\t/t_0)^{2/3}$ and the integration in~\eqref{lfs1} becomes straightforward. The free-streaming scale of the heavier daughter for $\t > \teq$ reads
\begin{equation}
\label{lfsdeltatauafter}
 \l_2^\text{fs} (\d, \t) = 179 \Mpc 
\left(\frac{\t}{10^{13} \seconds}\right)^\frac{1}{3}
\frac{(\d+1)^2 -1}{\d+1}
\left(
1- 0.0285 \left(\frac{\t}{10^{13} \seconds}\right)^\frac{1}{3}
\right) \, ,
\end{equation}
where now $\delta$ is smaller than one in any case. Inserting the minimal $\d$ from~\eqref{implicitlowboundondelta} we obtain the minimal free-streaming scale of the heavier daughter $(\l_2^\text{fs})_\text{min}$ for $\t > \teq$. It is depicted in Fig.~\ref{fig:lfsmin1}.

For decays occurring in the small window after $\tnrmin$ but before $\teq$ the curve in Fig.~\ref{fig:lfsmin1} represents a lower bound, because we did not consider that the Universe is radiation-dominated in this period, i.e., we used~\eqref{lfsdeltatauafter}. We find that a correction were small and thus not important for our purpose.
In any case, around this time the assumption of a sudden transition leads to a larger error in the estimate.

Lyman-$\a$ forest data constrain the scale of collisionless damping.
Constraints range between $\l_\text{fs} \lesssim 0.5 \Mpc$~\cite{Seljak:2006qw,Viel:2006kd}, tighter ones~\cite{Viel:2007mv} and those more relaxed due to the rejection of less reliable data~\cite{Abazajian:2005xn,Viel:2007mv,Boyarsky:2008xj}. We assume a constraint on the free-streaming scale $\l_\text{fs} \lesssim 1 \Mpc$ to apply, if all the dark matter originates from particle decay.
Since $(\l_2^\text{fs})_\text{min} \gg 1 \Mpc$
the bound~\eqref{nondom} from non-dominance with $\bmax=1$ is naive. The actual lower bound on $\d$ lies closer to the hot dark matter (HDM) bound of Sec.~\ref{sec:hdm} with $\bmax < 1$, cp.~\cite{Boyarsky:2008xj}. 
Consequently, the heavier daughter from a two-body decay producing the desired dark radiation cannot form the observed dark matter.
The HDM bound is the tightest bound obtainable taking into account the impact of the scenario on structure formation.

\paragraph{Minimal velocity today}
Using~\eqref{pofT} we can compute today's velocity $v^0$ of the decay products. Either a daughter is still relativistic, $p>m \Rightarrow v\simeq 1$, or has become non-relativistic, $p<m \Rightarrow v\simeq p/m$. With~\eqref{p1} and~\eqref{m2} we obtain for the heavier daughter
\begin{equation}
\label{v20}
 v_2^0 (\Tdec,\d) = \frac{T_0}{\Tdec} \frac{\phi_\text{rms}}{2} \frac{(\d+1)^2 - 1}{\d+1}  \left(\frac{\gastst}{\gastsd}\right)^\frac{1}{3} \, ,
\end{equation}
where $\phi_\text{rms}$
takes into account the exponential decay law in comparison to the sudden decay approximation
and is derived in Appendix~\ref{appendix:expdecay}. 
It is $\phi_\text{rms}=1$ if the decay occurs during radiation domination and $\phi_\text{rms}\simeq 1.09$ if the decay occurs during matter domination. 
Bounds on today's dark matter velocity assume a Fermi-Dirac distribution, while the decay products are distributed as derived in Appendix~\ref{appendix:expdecay}. While a transfer function is known~\cite{Kaplinghat:2005sy}, both distributions can be considered as having an equivalent effect on structure formation as long as they have an identical root-mean-square velocity~\cite{Jedamzik:2005sx}.
The minimal velocity of the heavier daughter today is obtained by inserting the lower bounds on $\d$ from Sec.~\ref{sec:delta-tau-plane} into~\eqref{v20}. We find
\begin{equation}
 (v_2^0)_\text{min} \simeq 15 \, \frac{\text{km}}{\text{s}} \, \frac{\phi_\text{rms}(\t)}{\phi_\text{rms}(\t \ll \teq)}   \D\Neff 
\left(\frac{0.1286}{\omegadm}\right)
\left(\frac{\gastst}{\gastsd}\right)^\frac{2}{3} .
\end{equation}
Like the minimal free-streaming scale for $\t<\teq$ it depends on the amount of dark radiation only.
Referring to~\cite{Jedamzik:2005sx} to suppress scales of the size of a galaxy cluster $v^0 \gtrsim 1$~km/s.
In this case bounds at least similar to the HDM constraints seem to apply. For decays later than $t_\text{BBN}$ we found a minimal velocity roughly 15 times larger and independent of the time of decay.
If $\gastsd\sim 228.75 \gg \gastst$ the velocity is significantly lowered but still too large to be compatible with Lyman-$\a$ limits.
This confirms that the non-dominance bound is naive and indeed HDM constraints apply in any case.

\subsubsection{Hot dark matter constraint and an opportunity}
\label{sec:hdm}

If a particle was relativistic at matter-radiation equality and became non-relativistic in the meantime, it represents hot dark matter (HDM).
In particular, observations of the structure in the Universe constrain
the amount of HDM, which counts towards $\O_\text{m}$ but does not form structures below its large free-streaming scale.
Observations are used to derive upper bounds on the sum of the masses of
the SM neutrinos, $\sum m_\nu$. Bounds vary between $0.4 \eV$ for the
minimal $\Lambda$CDM model including large scale structure data and $2.6
\eV$ if only CMB data are used and more free parameters are included~\cite{Hannestad:2010kz}.
We re-write such bounds as constraints on the HDM fraction $\O_\text{hdm}/\O_\text{dm} \leq \bmax$.
It is 
\begin{eqnarray}
\label{bmaxhdm}
 \bmax &=& \frac{\O_\text{hdm}^\text{max}}{\O_\text{dm}} = \left(\sum m_\nu\right)_\text{max} \frac{n_\nu^0}{\rhocrit} \O_\text{dm}^{-1} \nonumber \\
& =& 0.098 \left(\sum m_\nu\right)_\text{max}/ \eV \, ,
\end{eqnarray}
 where $\O_\text{dm} \simeq 0.21$ has been inserted and $n_\nu^0 = \frac{1}{11} \Neff^\text{SM} n_\g^0$ is today's number density of one neutrino species, if $n_\g^0$ denotes today's number density of CMB photons. 
We note that we have not taken into account the differing phase-space distributions of neutrino HDM, described by a Fermi-Dirac distribution, and decay products with a phase-space distribution given by exponential decay in an expanding universe, see Appendix~\ref{appendix:expdecay}. There are differences in the damping tails between decay-produced dark matter and warm dark matter~\cite{Kaplinghat:2005sy}. 
Nevertheless, given current measurement uncertainties and considering
that constraints on HDM arise from observations on larger scales we
assume the HDM bounds to apply without change.

As HDM the corresponding decay product were non-relativistic
today. Then its energy density is $\r_2= m_2 Y_2 s_0$, where $s_0$
denotes today's entropy density of the Universe. Its yield is the same
as the yield of the decaying particle if it did not decay, $Y_2=Y$. Thus
its energy density as HDM today is just suppressed by the
mass ratio of the two particles,
\begin{equation}
\label{O2hdm}
 \O_2^\text{hdm} h^2 = \frac{m_2}{m} \O h^2 = \frac{\O h^2}{\d+1} \, ,
\end{equation}
where $\O h^2$ is given by~\eqref{omegamother}.
Any constraint on the amount of HDM simply yields an implicit lower bound on $\d$ as in~\eqref{implicitlowboundondelta} with only $\O_\text{dm} \rightarrow \O_\text{hdm}^\text{max}$.
For the bound depicted in Fig.~\ref{fig:delta-tau-dr} we choose a tight constraint, $\sum m_\nu < 0.44 \eV$ at $95\%$ CL~\cite{Hamann:2010pw} corresponding to $\bmax=0.043$, because this yields a strong lower bound on $\d$.
The temperature when the heavier daughter becomes non-relativistic, corresponding to the obtained lower bound on $\d$ from HDM constraints for $\t< \tnrmin$, i.e., $T_2^\text{nr}(\d=\d_\text{min}^\text{hdm})$, is obtained using~\eqref{T2nrbound} with $\bmax$ given by~\eqref{bmaxhdm}.
 We see that $T_2^\text{nr}(\d=\d_\text{min}^\text{hdm})$ is lower than $\Teq$, which is to be expected. The bound appears self-consistent without further assumptions.

Stronger bounds on $\d$ can arise from observations that exclude the
heavier daughter becoming non-relativistic before a certain time. For
example, if CMB observations required the heavier daughter to become non-relativistic after photon decoupling, the actual lower bound would stem from this requirement. This possible constraint is depicted in Fig.~\ref{fig:delta-tau-dr} as well.

We point out an interesting fact that leads to an opportunity.
The neutrino mass scale sets two in principle independent quantities, i)
the neutrino energy density after they became non-relativistic, $\O_\nu
\simeq \sum m_\nu n_\nu$, and ii) the time at which neutrinos become non-relativistic. 
The CMB alone is not very much affected, if the neutrinos are still relativistic at the time of photon decoupling, cf.~Sec.~6.1 in~\cite{Komatsu:2008hk}. As the thermal origin and thus the neutrino temperature is understood, this restricts the CMB sensitivity to masses $m_\nu \gtrsim 0.6 \eV$.
The origin of HDM from a particle decay is very different. At what time the daughter becomes non-relativistic~\eqref{T2nr} depends on its mass only indirectly. The dependence is on the mass hierarchy to its mother as given by~\eqref{O2hdm}.
Interestingly, a heavier daughter with an energy density above the tight HDM bound from a decay with $\t\ll\teq$ becomes non-relativistic at a time before photons finally decouple, $t_2^\text{nr} < t_\g^\text{dpl}$.
In this case we expect the CMB to be sensitive to the heavier daughter
acting as HDM\@.
Such an amount of relativistic energy becoming non-relativistic during CMB times should leave observable consequences in the CMB even though the observational situation is not clear anymore.
For sure, observations on smaller scales are affected.

By considering galaxy cluster data corresponding to a scale of roughly
$10\Mpc$, $\Neff=3.91 \pm 0.42$ and $\sum m_\nu = (0.34 \pm 0.17) \eV$ (both 68\% CL) have been measured in~\cite{Benson:2011ut} assuming free $\Neff$ and $\sum m_\nu$. This deviates from zero by less than 2$\s$, but the maximum likelihood constraint is peaked away from zero. Improvements to these observations are said to be already approved. 
The obtained mean value might well be the first hint of the neutrino mass scale, while
 the largest mass-squared splitting from neutrino data is $\sqrt{|\D m_{31}^2|} \simeq 0.050 \eV$~\cite{GonzalezGarcia:2010er}.
In Fig.~\ref{fig:delta-tau-dr} the required $\d$ to have $\D\Neff=0.86$ from the lighter daughter and $\O_\text{hdm}=0.007$ corresponding to $\sum m_\nu =0.34 \eV$ from the heavier daughter corresponds exactly to the upper edge of the thick solid black curve.
Planck data combined with LSST or JDEM can constrain $\sum m_\nu < 0.04 \eV$~\cite{Abazajian:2011dt,Joudaki:2011nw}.
Experiments like KATRIN~\cite{Osipowicz:2001sq} or those seeking the neutrinoless double beta decay~\cite{Aalseth:2002rf,Arnold:2005rz} will measure neutrino masses in the laboratory.
We point out an interesting possibility given theses future sensitivities.
If laboratory experiments measure $m_\n$ smaller than cosmological  probes, this mismatch can be explained by a cosmological particle decay. The HDM contribution in the Universe could have originated from particle decay.
As we have shown the HDM and dark radiation can originate from the same decay.
Since the connection between observed mass scale and the time when the HDM becomes non-relativistic is different than for an additional relativistic species, these two cases should be distinguishable by future cosmological observations.

\section{Branching ratio constraints and opportunities}
\label{sec:brconstraints}
In this section we derive constraints on several branching ratios of the decaying particle.
Since the energy density of the decaying particle~\eqref{omegamother} is practically fixed for any significant value $\D\Neff>0$,
these bounds are general and, in particular, independent of the particle physics model.

\paragraph{Bounds from BBN}
\begin{figure}[t]
 \centering
   \includegraphics[width=0.9 \textwidth]{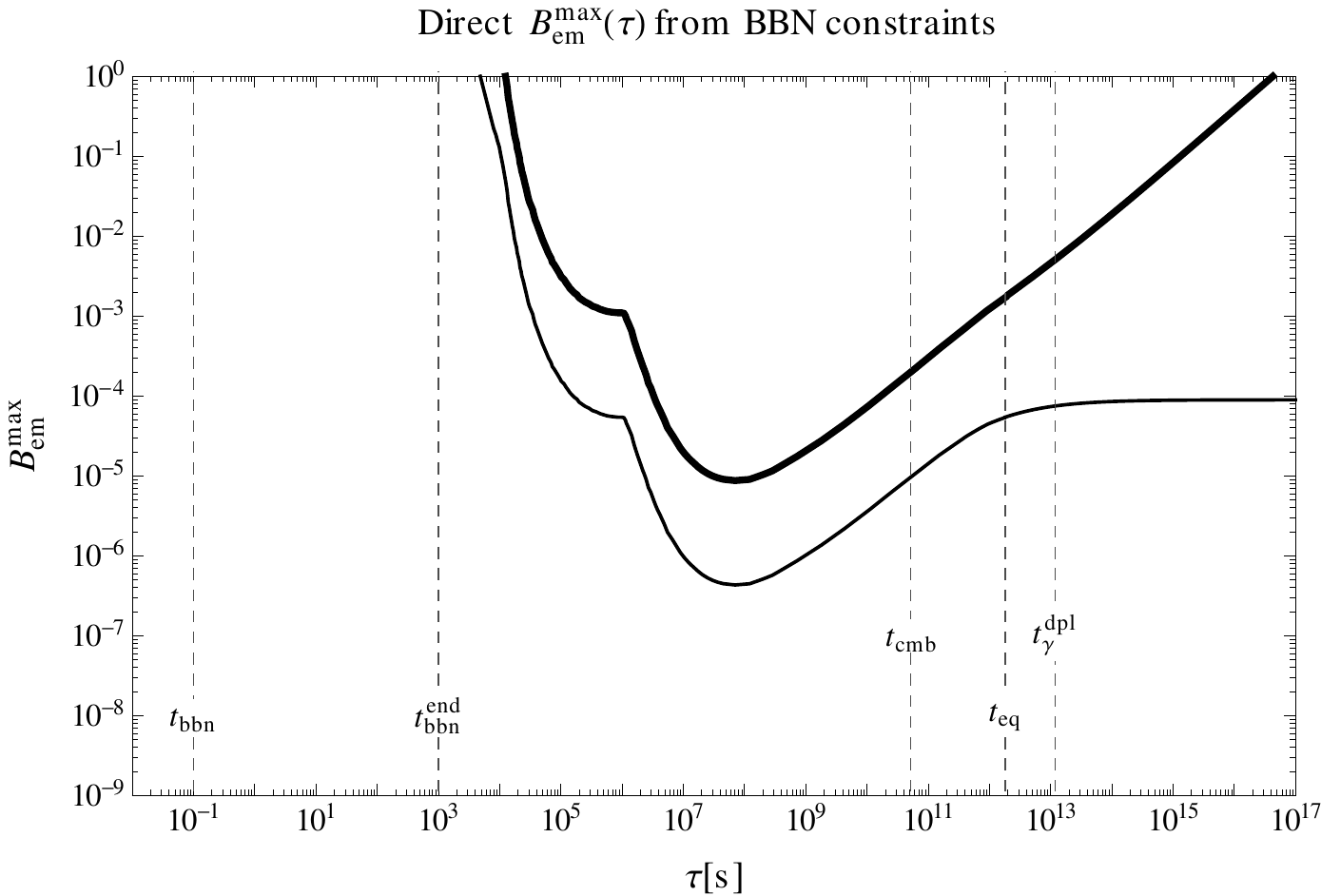}
\caption{Upper bound from BBN on the direct branching ratio of the decaying particle into electromagnetically interacting particles as function of its lifetime $\t$.
The thick solid curve represents the weakest bound obtainable. We consider larger branching ratios as excluded.
The thin solid curve represents the strongest bound obtainable. It shows how strong the actual bound might become depending on the actual value of $\D\Neff$ and the energy density of the decaying particle after $\teq$.
Times are highlighted as in Fig.~\ref{fig:delta-tau-dr}.
}
\label{fig:BemmaxBBNdirect}
 \end{figure}
We adopt the bounds from big bang nucleosynthesis (BBN) constraints determined in~\cite{Jedamzik:2006xz}
(see also \cite{Kawasaki:2004qu}).  They were derived assuming
the direct decay of some hypothetical massive particle into pairs of Standard Model particles, $X \rightarrow \text{SM} + \text{SM}$.
The product $\O \times B_\text{had/em}$ is bounded, where $B$ denotes
the branching ratio into either hadronically (had) or
electromagnetically (em) interacting pairs of Standard Model particles.
Electromagnetic primaries are, for example, photons, $\g\g$, and
electron-positron pairs, $e^+ e^-$. Hadronic primaries are quarks, $q
\bar q$, and gluons, $gg$. Hadronic decays generically also lead to the injection of electromagnetically interacting primaries, for example, due to neutral pions decaying into two photons.
The upper bound on the branching ratios is simply found as
\begin{equation}
 B_\text{had/em}^\text{max} = \frac{\O_\text{had/em}^\text{max}}{\O} \, ,
\end{equation}
where $\O_\text{had/em}^\text{max}$ is extracted from~\cite{Jedamzik:2006xz} and $\O$ is given by~\eqref{omegamother}.
We will argue that the bounds are indeed independent of the kind of coupling and the existence of only one dark decay mode. They do not depend on $\d$ either, because $\d$ is in any case bounded from below by some cosmological requirement. So they might depend on this cosmological bound but not on the particle physics parameters.

\begin{figure}[t]
 \centering
   \includegraphics[width=0.9 \textwidth]{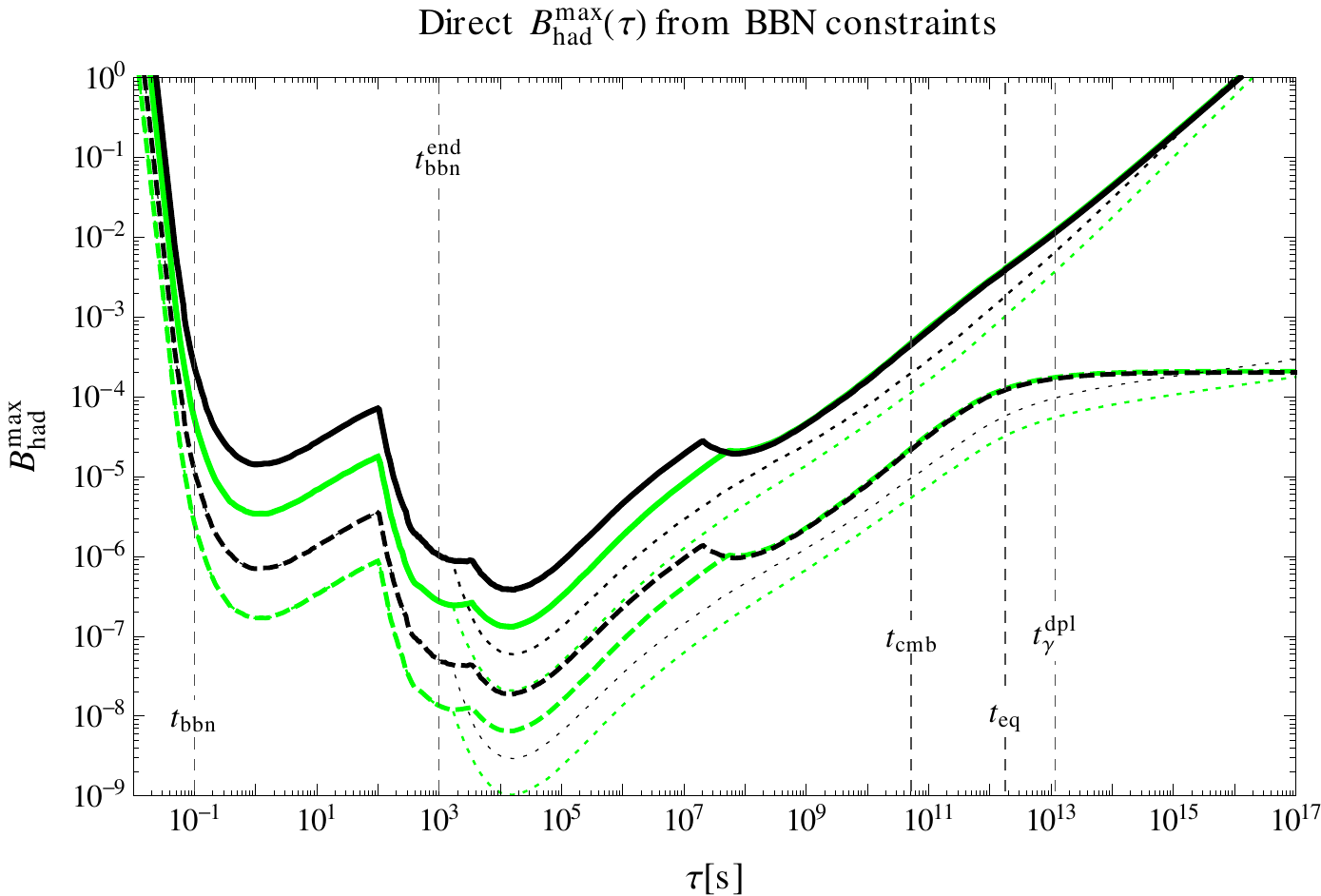}
\caption{Upper bound from BBN on the direct branching ratio of the decaying particle into hadronically interacting particles as function of its lifetime $\t$.
Solid/dashed curves represent weakest/strongest bounds obtainable as in Fig.~\ref{fig:BemmaxBBNdirect}.
Black curves apply to a mass of the decaying particle $m=1 \TeV$ and grey (green) curves to $m=100\GeV$.
The thin dotted curves correspond in each case to a less conservative bound for the ${}^6$Li$/{}^7$Li ratio.
Within the enclosed area the cosmic lithium problems could be solved by the decay.
Times are highlighted as in Fig.~\ref{fig:delta-tau-dr}.
}
\label{fig:BhadmaxBBNdirect}
 \end{figure}

Bounds from BBN depend on the cosmic time. The desired energy density of the decaying particle depends on the time of decay as well. Therefore, the resulting bounds are characteristic functions of the lifetime of the decaying particle. They are unique for the production of cosmic dark radiation.
Upper bounds from BBN on the direct branching ratio into electromagnetically interacting particles are depicted in Fig.~\ref{fig:BemmaxBBNdirect}.
Since they arise from the destruction of formerly built nuclei by photodisintegration, they become effective at rather late times $\sim 10^4 \seconds$. However, they become severe with the strongest bound, $B_\text{em}^\text{max} \sim 10^{-5}$, around $10^8 \seconds$. Afterwards the bound becomes weaker as the energy density of the decaying particle decreases following~\eqref{rhoatdec}.
Bounds are provided in~\cite{Jedamzik:2006xz} up to $10^{12} \seconds$. Towards later times we perform a trivial linear extrapolation. Such an extrapolation is crude, but we will see that CMB constraints are anyway stronger in this regime.
Since no $\D\Neff >0$ is confirmed, we have to consider a range of possible values. A minimal value $\D\Neff^\text{min}=0.52$ might be set by the expected 2-$\s$ exclusion limit of Planck~\cite{Perotto:2006rj,Hamann:2007sb}. As maximal value we take the 5-$\s$ exclusion of the combined analysis in~\cite{Dunkley:2010ge}, $\D\Neff^\text{max}=5.265$. These two different values lead to the spread of the two curves in Fig.~\ref{fig:BemmaxBBNdirect} before the time of matter-radiation equality $\teq$. This is the same for the solid and dashed curves in Fig.~\ref{fig:BhadmaxBBNdirect}.
The different behaviour after matter-radiation equality in both figures stems from the following: Before $\teq$ the conversion factor~\eqref{f} is always nearly $\mu$, because $\d$ must be larger than $\d_\text{min}$ given by~\eqref{lowboundondelta}. 
Here, we implemented the bound from the non-dominance requirement, because smaller $\d$ results at later times in a stronger strong bound.
After $\teq$ the maximal $\O$ is given by $\O_\text{dm}$ and
$\d_\text{min}$ becomes smaller than one, which leads to the plateau of
the strong bound after $\teq$. For the weakest bound obtainable the
conversion factor~\eqref{f} is $\mu$ also after $\teq$, because $\d$ can
also be arbitrarily large, if $\O \lesssim \O_\text{rad}$ instead. Thus the weak bound becomes weaker as  $\r_\text{rad}$  decreases relative to $\r_\text{mat}$ due to the expansion of the Universe.

\begin{figure}[t]
 \centering
   \includegraphics[width=0.9 \textwidth]{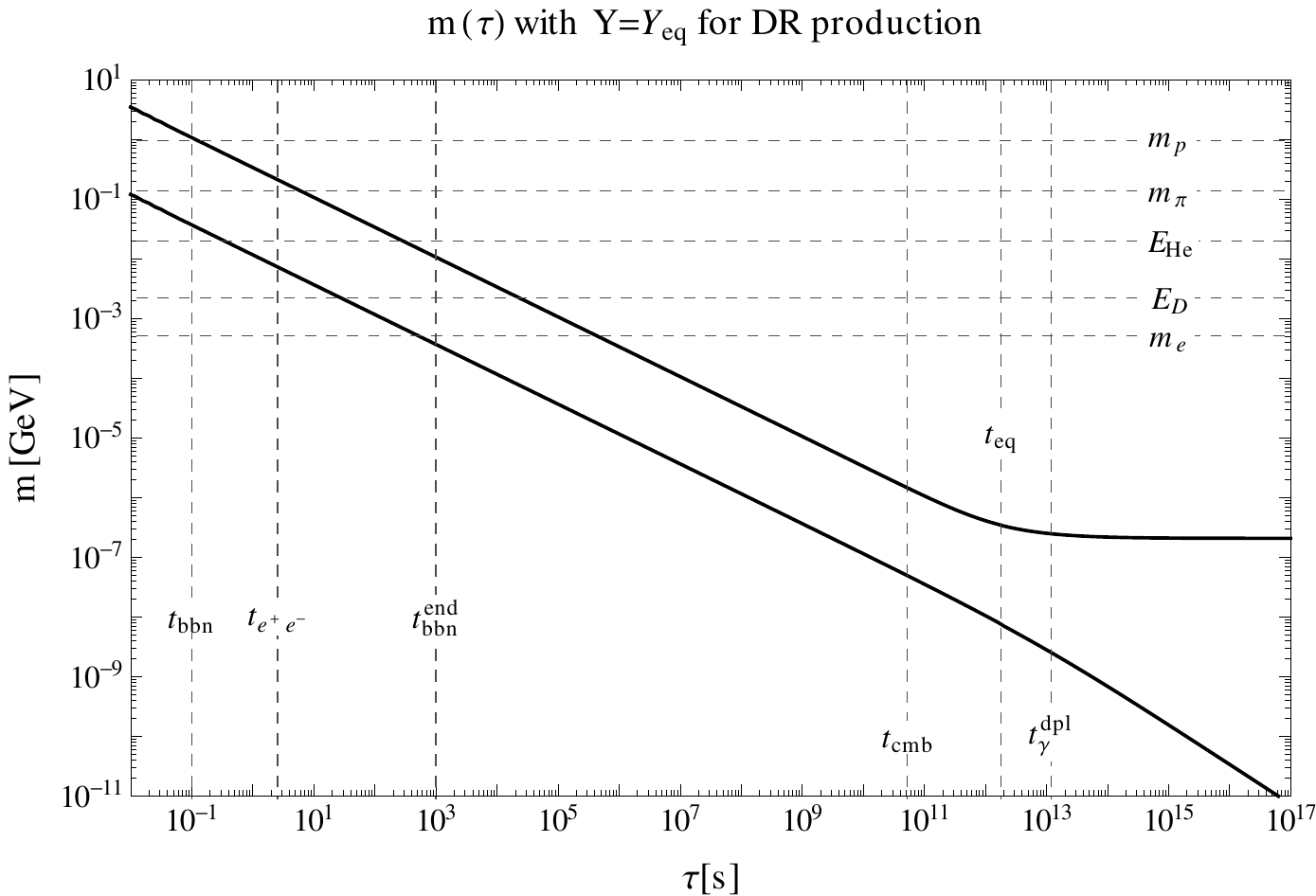}
\caption{Allowed masses $m$ for a decaying particle with equilibrium yield.
Within the lower ($mY=(mY)_\text{min}$, $Y_\text{eq}(\gast=\gast^\text{SM}=106.75)$) and upper ($mY=(mY)_\text{max}$, $Y_\text{eq}(\gast=228.75)$) line a decaying particle which left thermal equilibrium being relativistic at some early time might produce the desired amount of dark radiation depending on $\gast$ and the type of particle. 
Horizontal lines indicate masses of the proton $m_\text{p}$, the pion $m_\pi$, the electron $m_e$ and photodissociation thresholds of ${}^4$He and deuterium, respectively.
Times are highlighted as in Fig.~\ref{fig:delta-tau-dr}.
}
\label{fig:mYeq}
\end{figure}

Upper bounds from BBN on the direct branching ratio into hadronically interacting particles are depicted in Fig.~\ref{fig:BhadmaxBBNdirect}.
They arise from different processes at different times. Charged mesons and antinucleons affect relic abundances already at times as early as $ \sim 10^{-1} \seconds$. Therefore, the hadronic branching ratio is bounded, $B_\text{had}^\text{max} \sim 10^{-4}$--$10^{-6}$, already at such early times. The bound is strongest, $B_\text{had}^\text{max} \sim 10^{-6}$--$10^{-9}$,  around $ 10^4 \seconds$ and afterwards becomes weaker, because the energy density of the decaying particle decreases.
In determining the hadronic constraints we encounter specific additional uncertainties:
First, it has been shown in~\cite{Jedamzik:2006xz} that the hadronic BBN bounds depend not only on the time of decay, but also on the actual value of the hadronic branching ratio. 
We find that this dependence is too weak to be important for our purpose.
Therefore, we assume a pure scaling of the bounds with the branching ratio between the extremal cases $B_\text{had}=1$ and $B_\text{had}=0$.
Second, the bound on the hadronic branching ratio has also been shown to depend on the mass of the decaying particle.
If the mass is varied while the energy density is kept fixed, the number density varies accordingly. To first order different effects cancel out. Bounds for two different masses of the decaying particle are provided to indicate the remaining dependence. 
Black curves in Fig.~\ref{fig:BhadmaxBBNdirect} apply to a mass of the decaying particle $m=1 \TeV$ and grey (green) curves to $m=100\GeV$.
Referring to Fig.~\ref{fig:mYeq} we note that it might be motivated by the production of dark radiation from particle decay to extend an analysis of BBN constraints towards smaller masses, $m < 100 \GeV$, of the decaying particle.
Third, the determinations of ${}^6$Li and ${}^7$Li abundances are affected  by uncertainties in the understanding of nuclei destruction processes in stars.
The thin dotted curves in Fig.~\ref{fig:BhadmaxBBNdirect} correspond in each case to a less conservative bound for the ${}^6$Li$/{}^7$Li ratio. 
Branching ratios above these curves but below the corresponding more conservative bounds should not be regarded as ruled out.

In contrast, if they do not violate other bounds, they are a possible explanation for the relatively high ${}^6$Li/H ratios observed in metal-poor halo stars, providing the cosmic origin of ${}^6$Li~\cite{Jedamzik:1999di}. 
If the observationally inferred ${}^7$Li/H ratio is solved by stellar depletion, both problems, known as the cosmic lithium problems could be solved by the same particle decay.
At this point it is important to remind that the bounds in Fig.~\ref{fig:BemmaxBBNdirect} apply equally to hadronic primaries, because  they inject numerous electromagnetic primaries. These are also constrained in the following section.

\paragraph{Bounds from the CMB}
\begin{figure}[t]
 \centering
   \includegraphics[width=0.9 \textwidth]{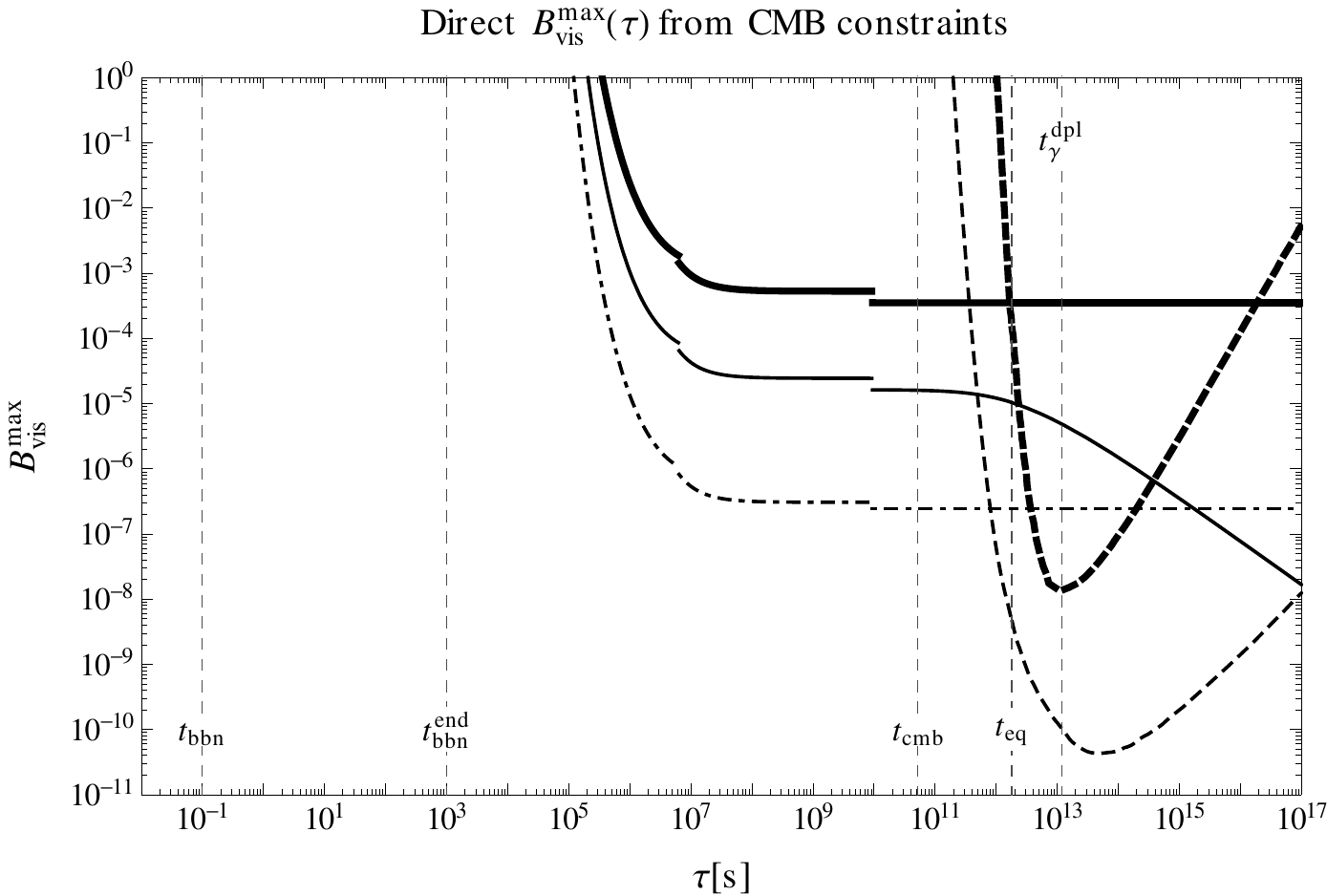}
\caption{Upper bound from the CMB on the direct branching ratio of the decaying particle into photons and electron-positron pairs as function of its lifetime $\t$.
Thick/thin solid curves represent weakest/strongest bounds obtainable as in Fig.~\ref{fig:BemmaxBBNdirect} from spectral distortions on the branching ratio into photons $B_\g$.
In the same sense thick/thin dashed curves represent weakest/strongest bounds obtainable from the ionisation history on the branching ratio into electrons, positrons and/or photons.
The dash-dotted curve indicates PIXIE's discovery reach, if $\D\Neff=1$, $Y_p=0.249$ and $\gdrobs=2$.
Times are highlighted as in Fig.~\ref{fig:delta-tau-dr}.
}
\label{fig:BphmaxCMBdirect}
 \end{figure}
The emission of particles with Standard Model interactions leads to spectral distortions of the cosmic microwave background (CMB). 
As in the previous section, we exploit the fact that the energy density of the decaying particle is fixed by the amount of dark radiation. 
To determine an upper bound on the branching ratio into photons we update the analysis of~\cite{Hu:1993gc} by taking into account the corrections pointed out in~\cite{Chluba:2011hw} and the current limits on deviations of the CMB from a thermal spectrum. These, obtained by COBE FIRAS, are $|\m| < 9 \times 10^{-5}$ and $y < 1.5 \times 10^{-5}$~\cite{Mather:1993ij,Fixsen:1996nj}.
 We find
\begin{equation}
\label{phbound}
 B_\g \lesssim 0.66 \,  \m_\text{max} \,\D\Neff^{-1} \gdrobs (\gastd)^\frac{5}{4} e^{(t_\m/\t)^\frac{5}{4}} \, ,
\end{equation}
where $\m_\text{max}$ denotes the bound on $\m$ and $t_\m \simeq 6.91 \times 10^6 \seconds \, (1-Y_p/2)^{4/5}$ the time scale of thermalisation in the Universe at that epoch. It is set by the primordial helium abundance $Y_p$ and other cosmological parameters, where we inserted PDG mean values~\cite{Beringer:1900zz}.
Following~\cite{Chluba:2011hw} we replace $e^{(t_\m/\t)^{5/4}} \rightarrow 0.48 (\t/t_\m)^{10/18} e^{1.99 (t_\m/\t)^{10/18}}$ for times earlier than $t_\m$.
 Upper bounds on the direct branching ratio into photons $B_\g$ from spectral distortions of the CMB are depicted as solid curves in Fig.~\ref{fig:BphmaxCMBdirect}.
They are not effective at times earlier than $3 \times 10^5 \seconds$,
because injected photons thermalise safely, not leaving any observable imprint. In the analytic approximation~\eqref{phbound} this is explicated by the exponential factor.
For $\t \gtrsim t_\m$ the weak bound becomes constant on a severe level, $B_\g^\text{max} \simeq 4 \times 10^{-4}$. This is a qualitative difference to the BBN bounds.
At a certain time around $2\times 10^{11} \seconds$ the CMB constraints become stronger than the bounds in Fig.~\ref{fig:BemmaxBBNdirect}. Therefore, the extrapolation of the BBN bounds towards later times is not crucial for our purposes.
The spread between the strongest bounds obtainable (thin curves) and the weakest bounds obtainable (thick curves) is mainly due to the same reasons as for the bounds from BBN on the electromagnetic branching ratio depicted in Fig.~\ref{fig:BemmaxBBNdirect}.
To consider the relatively large observational uncertainty in the primordial ${}^4$He abundance we take for the strong bound $Y_p=0.267$ and for the weak one $Y_p=0.231$, which corresponds to the 2-$\s$ statistical and systematic PDG error range~\cite{Beringer:1900zz}.
No change of slope arises at $\teq$ from the change in the expansion law, because it turns out that in our parametrisation  the bound function~\eqref{phbound} has no proportionality to the time of decay.
\begin{figure}[t]
 \centering
   \includegraphics[width=0.9 \textwidth]{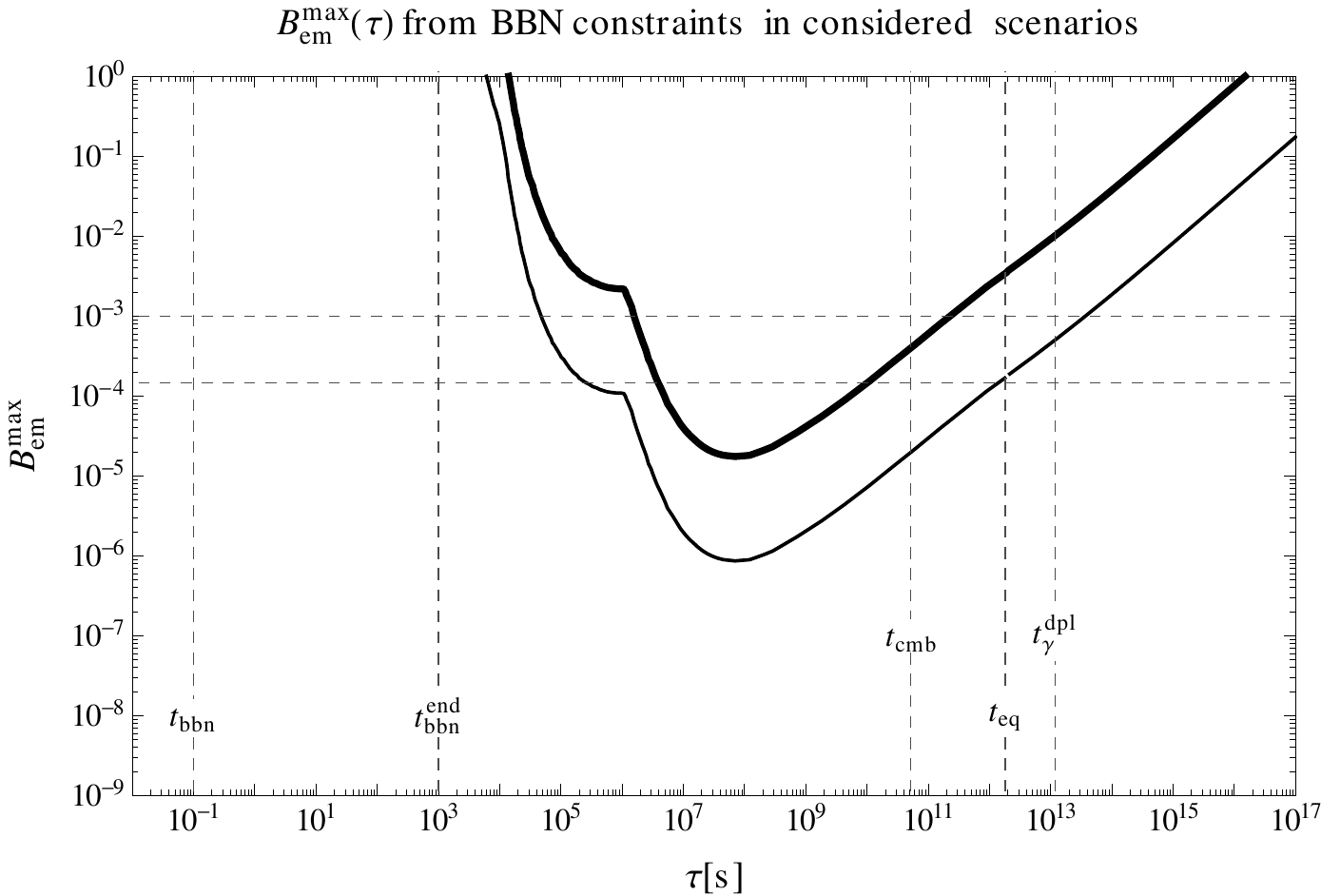}
\caption{
As Fig.~\ref{fig:BemmaxBBNdirect} but on the off-shell or non-tree-level branching ratio of the decaying particle as applying to the scenarios under consideration.
Horizontal lines indicate i) the often quoted suppression of $10^{-3}$ for off-shell processes and ii) the suppression from an electromagnetic loop and three-to-two-body kinematics. 
}
\label{fig:BemmaxBBN}
\end{figure}
For times $\t \lesssim 4 \, \omegab \times 10^{11} \seconds$ elastic Compton scattering establishes a Bose-Einstein spectrum with chemical potential $\m$ regardless of the details of the injection.
For the considered case of massive particle decay, the number of injected photons is negligible relative to the number of photons in the background.
In any case the energy density of injected photons has to be small compared to the energy density of background photons. Then the induced chemical potential is proportional to the injected energy density, $\m \propto B_\g \r/\r_\g$.
For times $\t \gtrsim 4 \omegab \times 10^{11} \seconds$ the spectrum can be described by the Compton $y$-parameter.
The jump in the bound due to this change in the description of the spectrum with the corresponding constraints is easily identified in Fig.~\ref{fig:BphmaxCMBdirect} around $10^{10} \seconds$.
It is $B_\g \r/\r_\g = 4 y$.

\begin{figure}[t]
 \centering
   \includegraphics[width=0.9 \textwidth]{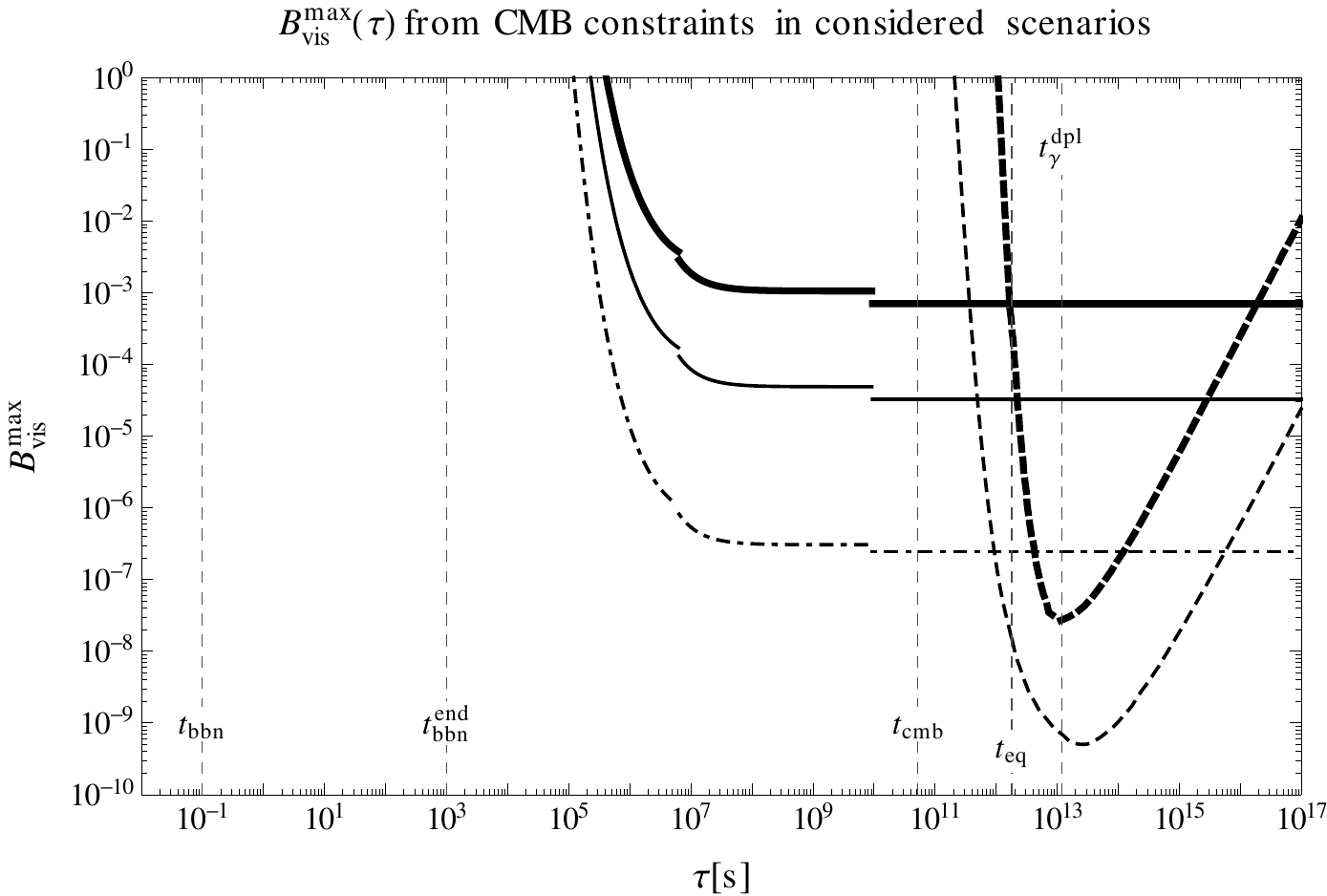}
\caption{
As Fig.~\ref{fig:BphmaxCMBdirect} but on the off-shell or non-tree-level branching ratio of the decaying particle as applying to the scenarios under consideration.
}
\label{fig:BphmaxCMB}
\end{figure}

Future CMB polarimeters such as PIXIE are proposed providing dramatically tighter constraints with projected detection levels of $\m \sim 5 \times 10^{-8}$ and $y \sim 10^{-8}$ at 5-$\s$~\cite{Kogut:2011xw}. 
The dash-dotted curve indicates PIXIE's discovery reach, if $\D\Neff=1$, $Y_p=0.249$ and $\gdrobs=2$.
PIXIE could even identify the origin of a spectral distortion as particle decay~\cite{Chluba:2011hw,Khatri:2012tw}.
Comparing Figs.~\ref{fig:BphmaxCMBdirect} and~\ref{fig:BphmaxCMB} with the corresponding hadronic bounds we see that due to the specific behaviour of the bounds large portions of parameter space probed by PIXIE are neither excluded by BBN nor by changing  the ionisation history of the Universe.
At times $\t > \teq$ the mother could form structures, in particular, if
its energy density is large, $\Omega \simeq \Omega_\text{dm}$.
Its decay should then lead to inhomogeneous $\m$-distortions~\cite{Pajer:2012vz},
which could allow to derive stronger bounds than the one from
homogeneous distortions. However, as we shall see, bounds from changes
of the ionisation history are severe at such late times and typically
much stronger than the bound from homogeneous $\m$-distortions.
Therefore, we do not elaborate on this possibility.

The emission of particles with Standard Model interactions may change
the ionisation history of the Universe, which can leave observable
consequences in the CMB\@.
We adopt the bounds on scenarios with late-decaying particles derived in~\cite{Slatyer:2012yq} based upon WMAP7 limits.
 Upper bounds on the direct branching ratio into electron-positron pairs or photons from additional ionisation and heating observable in the CMB are depicted as dashed curves in Fig.~\ref{fig:BphmaxCMBdirect}.
They become effective at a much later time around $\teq$, but also quickly stronger than the bounds from spectral distortions.
Towards times $\t <10^{12}\seconds$ we perform an extrapolation using a linear and quadratic term.
 They are strongest already around $t_\g^\text{dpl}$ and then become weaker towards later times as the energy density of the decaying particle decreases and a decay has less impact on observables. Nevertheless, they stay strong till today, in particular, stronger than bounds from spectral distortions at late times.
In~\cite{Slatyer:2012yq} for each lifetime it is scanned over the constraint for photons and electron-positron pairs for masses of the decaying particle ranging from $2 \keV$ to $12 \TeV$. The bound becomes a band with its width reflecting the variation between different decaying particle masses and decay products.
To consider this uncertainty we take the strongest bound provided for our strong bound and the weakest bound provided for our weak bound, respectively. This approach is sufficient for our purpose, because the weakest bound obtainable from spectral distortions is already quite strong.
Nevertheless, we would like to point out that this uncertainty --for
example, $1.1$ orders of magnitude at $\t =10^{14} \seconds$-- could be
reduced to a negligible level referring to~\cite{Slatyer:2012yq}. Since
we provide the injection spectrum and the variation with red-shift, one
should be able to do so using the provided grid of injection energies
and red-shifts.
Referring to Fig.~\ref{fig:mYeq} we note that it might be motivated by the production of dark radiation from particle decay to extend the analysis of these constraints towards smaller masses, $m < 1 \keV$, of the decaying particle.
\begin{figure}[t]
 \centering
   \includegraphics[width=0.9 \textwidth]{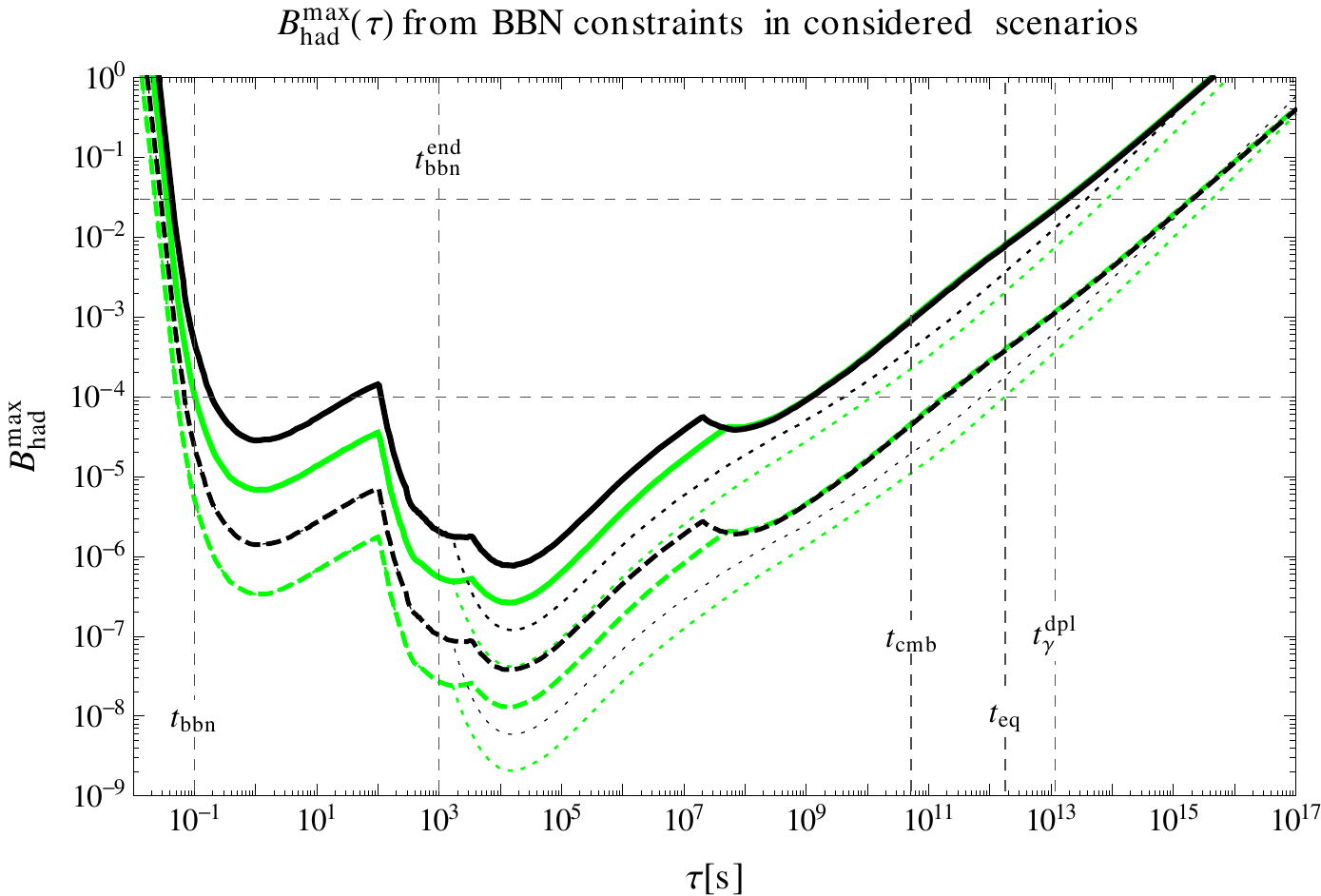}
\caption{
As Fig.~\ref{fig:BhadmaxBBNdirect} but on the off-shell or non-tree-level branching ratio of the decaying particle as applying to the scenarios under consideration.
Horizontal dashed lines indicate: i) the minimal hadronic branching ratio $\simeq 0.03$ of a neutralino decay into gravitino calculated in~\cite{Covi:2009bk} and 
ii) the minimal hadronic branching ratio $\sim 10^{-4}$ of a sneutrino decay into gravitino for sneutrino masses $\gtrsim 200 \GeV$ and $\d>9$ found in~\cite{Kanzaki:2006hm}. 
}
\label{fig:BhadmaxBBN}
\end{figure}

\paragraph{Bounds on considered scenarios}
In the previous paragraphs we determined cosmological constraints on the direct decay, $X \rightarrow \text{SM} + \text{SM}$, into various decay products with Standard Model interactions.
As discussed these bounds are severe and any model must satisfy them.
For example, if photons are emitted at tree-level, the decay
has to occur before $\t \lesssim 10^4 \seconds$ and so on.
In Sec.~\ref{sec:oddms} we considered models without any tree-level
decay mode that could give rise to a constrained branching ratio. Dangerous terms
in the Lagrangian are forbidden by kinematics or by symmetries, which is relatively simple. This
means that the constrained direct branching ratios are automatically
zero. Thus, all constraints are fulfilled by construction. 
For the models studied in Sec.~\ref{sec:tddms}, we assume that
either this holds true as well or that dangerous tree-level modes are sufficiently suppressed. We repeat that this often means branching ratios smaller than $10^{-4}$.
In any case decays emitting particles with SM interactions may proceed
off-shell (e.g., $X \to \text{dark} + \text{?}^\ast \to \text{dark} + \text{SM} + \text{SM}$)
or via loop processes. Then the previously given bounds do not apply.  Three-body final states might yield the leading contribution.
In short, we have to consider how much energy is carried away invisibly in each decay.

At early decay times, $\t \ll \teq$, the situation is very simple. Half
of the energy will always be carried away by some dark decay product as
we found already that both daughters have to be emitted relativistically
with equal momenta. The case that both decay products go off-shell
can safely be neglected, because the matrix element is additionally suppressed and because four-body final states are also suppressed kinematically.
Likewise, loop-induced three-body decays are suppressed relative to a
loop-induced two-body decay and therefore negligible.
At later decay times, $\t > \teq$, heavier decay products might be
emitted non-relativistically. Actually, this is by construction the case
for the strong bounds depicted in
Figs.~\ref{fig:BemmaxBBNdirect},~\ref{fig:BhadmaxBBNdirect}
and~\ref{fig:BphmaxCMBdirect}. The virtuality of particles with $p<m$ is
small. It is very improbable that they decay off-shell.
Therefore, it is a very good approximation to assume them to be safe. Only the decay products forming the dark radiation or --at least-- being emitted relativistically are left to endanger observations in this case.
Altogether, to determine upper bounds on the off-shell or non-tree-level branching ratio of the decaying particle we multiply~\eqref{omegamother} by $((\d+1)^2 -1)/(2(\d+1)^2)$. This is to consider the amount of energy which is always carried away invisibly. The factor cancels the dependence on $\d$ in~\eqref{omegamother}.
Actually,
it is universal in the sense of quantifying the relativistically emitted fraction of energy as we found that $\d$ is not free, but has to satisfy bounds. Applied to the scenarios of Sec.~\ref{sec:tddms} this means $B_\text{dr}= B_1 = \frac{(\d_\text{min}+1)^2 -1}{(\d_\text{min}+1)^2}$, while the situation is odd, because it is restricted to a particular window of mass hierarchies, $1/4 < x_2  < 1/2$, see Sec.~\ref{sec:tddms}.
The resulting bounds are shown in
Figs.~\ref{fig:BemmaxBBN},~\ref{fig:BhadmaxBBN} and~\ref{fig:BphmaxCMB}
analogously to Figs.~\ref{fig:BemmaxBBNdirect},~\ref{fig:BhadmaxBBNdirect} and~\ref{fig:BphmaxCMBdirect}.
Compared to those on direct branching ratios all bounds are reduced by a
factor $1/2$. More importantly, there is no additional spread between strong and weak bound after $\teq$ for off-shell and non-tree-level branching ratios.
 
The obtained bounds are severe and have the power to exclude many
particle physics possibilities. As a demonstration, we give two examples.
i) According to Fig.~\ref{fig:BemmaxBBNdirect} a neutralino LOSP could decay during or even after BBN, $\t \lesssim 10^4 \seconds$, into a
gravitino and a photon. The photon thermalises safely and the gravitino could act as dark radiation. However, as shown in Fig.~\ref{fig:BhadmaxBBN} this is excluded for $  4 \times 10^{-2} \seconds \lesssim \t \lesssim t_\g^\text{dpl}$ due to the minimal hadronic branching ratio of this decay. It arises from the off-shell decay of the massless photon into $q \bar q$ pairs enhanced by a logarithmic infrared divergence, see~\cite{Covi:2009bk} for details. It is thus too large even for neutralino masses as small as some tens of GeV\@.
ii) A sneutrino LOSP can decay invisibly into a gravitino and a
neutrino, both possibly acting as dark radiation. The bounds in
Fig.~\ref{fig:BemmaxBBNdirect},~\ref{fig:BhadmaxBBNdirect}
and~\ref{fig:BphmaxCMBdirect} are evaded. However, the weak interactions
of the sneutrino and the neutrino lead to a hadronic branching ratio
with significant impact on BBN, if the sneutrino is sufficiently heavier
than the electroweak gauge bosons. The minimal branching ratio of about $10^{-4}$ for sneutrino masses $\gtrsim 200 \GeV$ and $\d>9$ found in~\cite{Kanzaki:2006hm} is depicted in Fig.~\ref{fig:BhadmaxBBN}. We see that the decay is excluded for
$  10^{-1} \seconds \lesssim \t \lesssim 10^9 \seconds$.

In Fig.~\ref{fig:mYeq} we show the band of allowed masses for a decaying
particle that entered thermal equilibrium and then decoupled being relativistic at some early time. Thus it has a relatively large yield. Larger yields can be reached in non-thermal production mechanisms. 
The figure serves to make an important point. We can see that for such
yields the mass of the mother is restricted to be smaller than about $1 \GeV$ for decays occurring during or after BBN\@. For masses of the mother smaller than the proton mass no nuclei can be emitted. Thus no BBN constraints from the injection of nuclei apply. For masses of the mother smaller than the pion mass no hadronically interacting particles can be emitted at all and so on. Emitted photons with energies below photodissociation thresholds do not destroy nuclei. 
A small enough mass of the decaying particle can thus circumvent all BBN constraints. 
It might be motivated by the production of  dark radiation from particle  decay to extend analyses of BBN constraints towards masses smaller than $100 \GeV$.
Regarding the ionisation history of the Universe, the effect of photons
emitted with energies smaller than one keV seems worth studying as well.

\section{Two dark decay modes}
\label{sec:tddms}
In this section we study the origin of dark radiation from the two-body decay of a non-relativistic particle, which possesses two dark decay modes with corresponding branching ratios $B_{1,2}$ summing up to roughly one,
\begin{equation}
\label{b1b2}
B(X \rightarrow 1 +1) + B(X \rightarrow 2 +2) = B_1 +B_2 \simeq 1 \, ,
\end{equation}
where $1$ and $2$ denote and label the two dark decay products.
We showed in Sec.~\ref{sec:brconstraints} that the branching ratio into dark components is constrained to be very close to one at times later than $t_\text{BBN}$.
Considering the upper right corner of Fig.~\ref{fig:energydensities},
there is no heavier and lighter daughter in each decay now. In contrast, there are lighter daughters from one decay mode and heavier daughters from an additional decay mode.
Compared to the cosmologies in Sec.~\ref{sec:oddms} there is one additional parameter, the relative branching $B_1/B_2$.
This allows for the dash-dotted curve in Fig.~\ref{fig:energydensities}, where $B_2$ is obviously much smaller than one.

Well-motivated examples for such decays are i) saxion decays into two axinos with $y^2 \sim \msax^2/(4f_a^2)$, where $\msax$ denotes the saxion mass and $f_a$ the axion decay constant, ii) moduli decays into two gravitinos with $y^2 \sim \k^2 m_\phi^2/(18\mplanck^2)$, where $m_\phi$ denotes the modulus mass and $\k$ an effective coupling,
iii) saxion decays into two axions with $y^2 \sim x^2 \msax^2/(4f_a^2) $, iv) flaton decays into two axions with $y^2 \sim m_f^2/(2f_a^2) $~\cite{Chun:2000jr}, where $m_f$ denotes the flaton mass or v) moduli decays into bulk axions with $y^2 \sim m_\phi^2/(2\mplanck^2)$~\cite{Cicoli:2012aq,Higaki:2012ar}. In i) and ii) a scalar decays into two fermions, while in iii)--v) a scalar decays into two scalars. Other combinations of spins are imaginable.
Note that in theories beyond the Standard Model with a dark matter stabilising symmetry either interactions as thought about  here  or in Sec.~\ref{sec:oddms} may be allowed for one and the same particle. Depending on the number of long-lived particles in the theory there may be particles of each kind in the spectrum. As one might infer from the given lists, this is the case, for example, in supergravity theories amended by the Peccei-Quinn mechanism.
A large number of long-lived particles may appear in theories with a
large dark sector, such as dynamical dark matter~\cite{Dienes:2011ja}.
One can imagine many scenarios where several particles decay with
different lifetimes and different numbers of dark decay modes,
potentially even more than two.  This includes the possibility of
dark cascades.

Even though all decay products may act as dark radiation, especially, the heavier ones do not need to do so at any time. 
In contrast, they may form the observed dark matter. Then two of three
dark components originate from the decay of the same particle.
This was proposed first in~\cite{Choi:2012zna}, where the authors also take into account BBN constraints.

\subsection{Basics}
We introduce useful parameters, determine general properties and derive basic equations.

\paragraph{Kinematics}
In each decay the two decay products have equal mass, so that the general expression for the initial momentum~\eqref{genmomentum} of each emitted particle reduces to
\begin{equation}
\label{pini2}
 p_\text{ini}= p_{1,2}= \frac{m}{2} \left( 1 - 4 x_{1,2}^2 \right)^\frac{1}{2}
\end{equation}
with $x_{1,2}= m_{1,2} /m$ denoting the mass ratio of the corresponding
daughters with equal mass and the mother. This  parameter is a
useful measure of their mass hierarchy. We choose labels such that $m_1
< m_2$. We have always $m_1, \, m_2 < m/2$, since otherwise the corresponding decay were kinematically forbidden. 
Non-relativistic emission can occur only in the mass window $ m/4 < m_2  < m/2$.
This scenario might appear unattractive for model building from the naturalness point of view.
The other way around, a theory predicting $m_2$ in this range would interestingly always lead to this case.

\paragraph[On Tnr and the 1/x-tau plane]{On $T_{1,2}^\text{nr}$ and the $1/x$-$\t$ plane}
\begin{figure}[t]
 \centering
   \includegraphics[width=0.9 \textwidth]{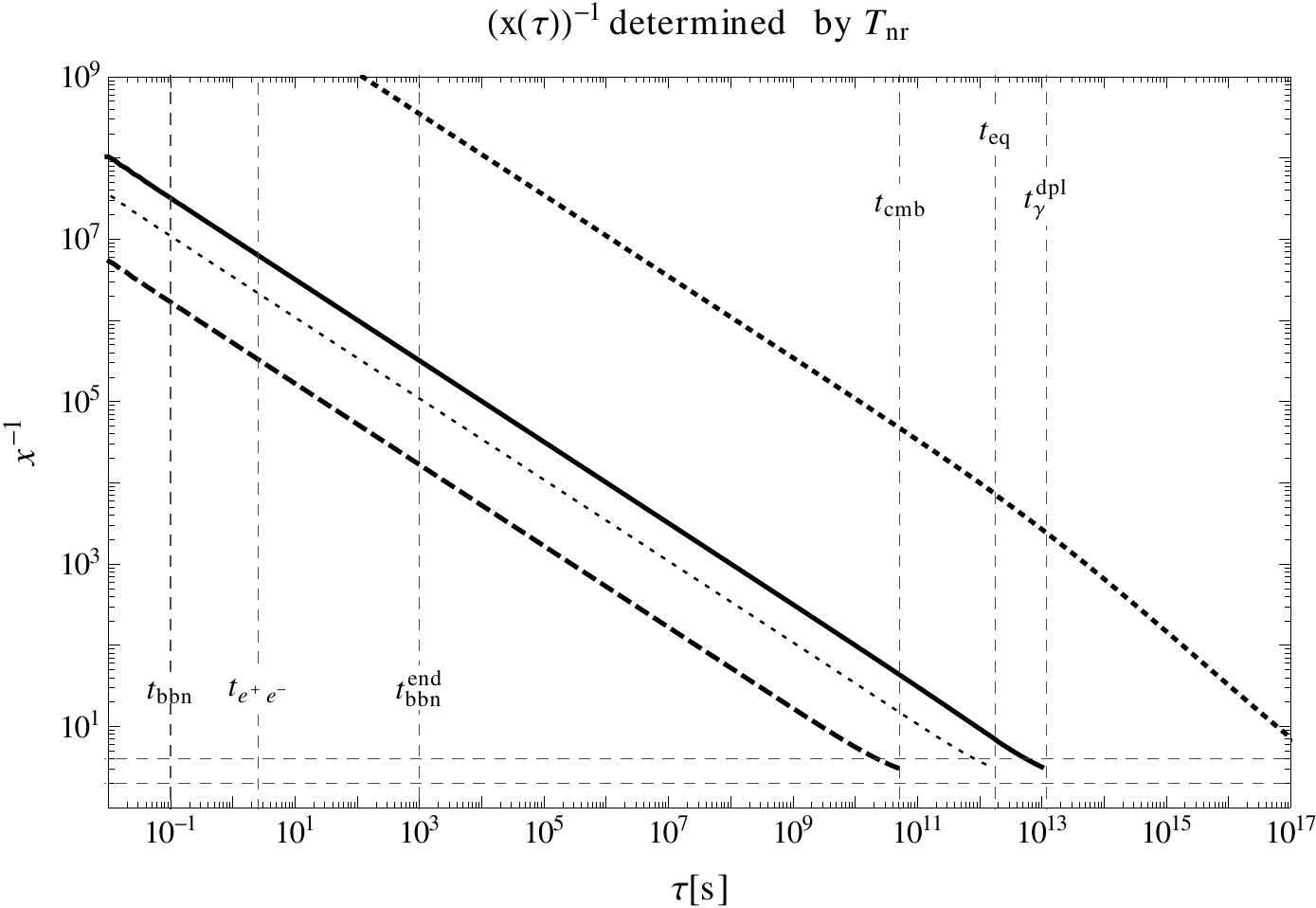}
\caption{Mass hierarchy $1/x_{1,2}=m/m_{1,2}$ given by~\eqref{x12} for different requirements on $T_{1,2}^\text{nr}$.
On the solid curve the decay products become non-relativistic at the time of photon decoupling. For smaller values of $x$ particles become non-relativistic at later times.
In the same sense the dashed curve corresponds to $t_{1,2}^\text{nr}=\taumax$, the thin dotted curve to  $t_{1,2}^\text{nr}=\teq$ and for a decay product to be still relativistic today $1/x_{1,2}$ would have to lie above the thick dotted curve.
Within the horizontal dashed lines the corresponding decay product is
emitted non-relativistically. Values of $x>1/2$ are not possible.
Times are highlighted as in Fig.~\ref{fig:delta-tau-dr}.
}
\label{fig:xmin}
\end{figure}
From the same general condition~\eqref{nrcond} we find the temperature when the corresponding daughters become non-relativistic, 
\begin{equation}
\label{Tnr12}
 T_{1,2}^\text{nr} = \Tdec \frac{2}{\m} 
\left( x_{1,2}^{-2} -4        \right)^{-\frac{1}{2}}
\left(\frac{\gastsd}{\gastsnr}\right)^\frac{1}{3} 
\end{equation}
with $\gastsnr=\gasts(T_{1,2}^\text{nr})$.
Typical requirements on $T_{1,2}^\text{nr}$ might be that i) the particle is still relativistic today, $T_{1,2}^\text{nr} < T_0$, which is the best --or only-- understood situation from the observational point of view, ii) the particle does not become non-relativistic during CMB times, $T_{1,2}^\text{nr}< T_\g^\text{dpl}$ or $T^\text{nr}_{1,2} > T(\taumax)$, which would possibly leave some observable signature in the CMB, or iii) the particle does or does not act  as radiation at matter-radiation equality, $T_{1,2}^\text{nr}< \Teq$ or $T_{1,2}^\text{nr}> \Teq $.
Such requirements turn into constraints on $x_{1,2}$ depending on the time of decay. From~\eqref{Tnr12} it is straightforward to single out
\begin{equation}
\label{x12}
 x_{1,2} = \frac{1}{2} 
\left(
\left(\frac{\Tdec}{T_{1,2}^\text{nr}}\right)^2 
\frac{4}{\m^2} 
\left(\frac{\gastsd}{\gastsnr}\right)^\frac{2}{3} + 4
\right)^{-\frac{1}{2}} .
\end{equation}
The constraints on $x_{1,2}$ from the requirements i)--iii) on $T_{1,2}^\text{nr}$ are shown in Fig.~\ref{fig:xmin}. 

\paragraph{Energy densities}
The desired amount of dark radiation determines the energy density of the decaying particle.
In principle, both decay channels can contribute to an increase in
$\D\Neff$, which might be interesting on its own only if both add some
observable contribution, i.e., for $B_1 \sim B_2$. There is a plethora of possible cosmologies.
For example, both channels can contribute dark radiation at BBN and/or photon decoupling. Heavier decay products could become non-relativistic in the meantime, while lighter ones are still relativistic today, and so on.
We shall focus on the case that the dark radiation at the time of
observation is formed by the lighter daughters only. However, the
following holds analogously in and can be applied to other cases as well.

Since $B_1+B_2 \simeq 1$, the number densities of the decay products at decay are given by
\begin{equation}
\label{n12}
 n_{1,2} \simeq 2 n B_{1,2} \, .
\end{equation}
If the dark radiation is formed by the lighter daughter particles,
\begin{equation}
 \rhodr = \r_1 =  n_1 E_1 \, , 
\end{equation}
where $E_1 \simeq \langle p_1 \rangle$, because they must be relativistic to act as radiation.
At decay $p_1$ is given by~\eqref{pini2} and $n_1$ by~\eqref{n12}. In this way the energy density of the decaying particle $\r = nm$ at decay is set by the amount of dark radiation. 
The mean $\langle p_1 \rangle/p_\text{ini} = \m$ is given by the distribution in Appendix~\ref{appendix:expdecay}.
We obtain
\begin{equation}
\label{rhodec2}
 \r|_\text{dec} = \m^{-1} B_1^{-1} \left( 1- 4 x_1^2 \right)^{-\frac{1}{2}} \rhodr(\Tdec) \, , 
\end{equation}
where $\rhodr(\Tdec)$ was found in~\eqref{rhodr}.
The conversion factor defined in~\eqref{deff} reads now
$
f = \m B_1 (1-4 x_1^2)^{1/2}
$.

After the heavier daughters become non-relativistic, their energy
density $\r_2 = n_2 E_2 \simeq n_2 m_2$ can be related to the energy
density of the mother, if it had not decayed, by making use of the definition of $x_2$ and inserting~\eqref{n12}. It is
$
 \r_2 = 2 B_2 x_2 \r  
$
or today
\begin{equation}
\label{omega22}
 \O_2 h^2 = 2 B_2 x_2 \O h^2 = 2 B_2 x_2 
\left(\frac{T_0}{\Tdec}\right)^3
\frac{\gastst}{\gastsd}
\frac{\r|_\text{dec} h^2}{\rhocrit} \, ,
\end{equation}
where the energy density at decay is fixed by~\eqref{rhodec2}.
Note that this is valid only for times when the heavier daughters are non-relativistic.

\subsection{Non-dominance requirement and dark matter}
The non-dominance requirement, cp.~\eqref{nondom}, on the heavier daughters gives by~\eqref{omega22} rise to an upper bound
\begin{equation}
\label{nondomb2x2}
 \frac{x_2}{B_2^{-1}-1} \leq 2.710 \times 10^{-3} \, \m \left(\frac{1\keV}{\Tdec}\right)
\left(\frac{\gastst}{\gastsd}\right)^\frac{1}{3}
\D\Neff^{-1} \left(\frac{\omegadm}{0.1286}\right) 
\left(1 - 4 x_1^2 \right)^\frac{1}{2} ,
\end{equation}
where we used furthermore that
$B_2/B_1 = 1/(B_2^{-1}-1)$ due to~\eqref{b1b2}.
For $\t > \teq$ the heavier daughters are either emitted
relativistically like the lighter ones or their mass 
is restricted to the window $m/4 <  m_2 < m/2$.  For $\t< \teq$ the requirement simplifies because i) $x_1$ is much smaller than one, cp.~Fig.~\ref{fig:xmin}, and
ii) $B_2$ will be typically much smaller than one for the heavier daughter to make up a viable dark matter candidate, so $(B_2^{-1}-1)^{-1} \simeq B_2$. Then~\eqref{nondomb2x2} reduces to
\begin{equation} \label{nondomb2x2approx}
 B_2 x_2 \lesssim 6.621 \times 10^{-3}
\left(\frac{\t}{10^7 \seconds}\right)^\frac{1}{2}
\D\Neff^{-1} \left(\frac{\omegadm}{0.1286}\right)
\left(\frac{\gastd}{\gastt}\right)^\frac{1}{4}
\left(\frac{\gastst}{\gastsd}\right)^\frac{1}{3} .
\end{equation}
A theory linking $B_2 x_2$ such that this non-trivial constraint is fulfilled naturally appears particularly attractive.

\begin{figure}[!ht]
 \centering
   \includegraphics[width=0.9 \textwidth]{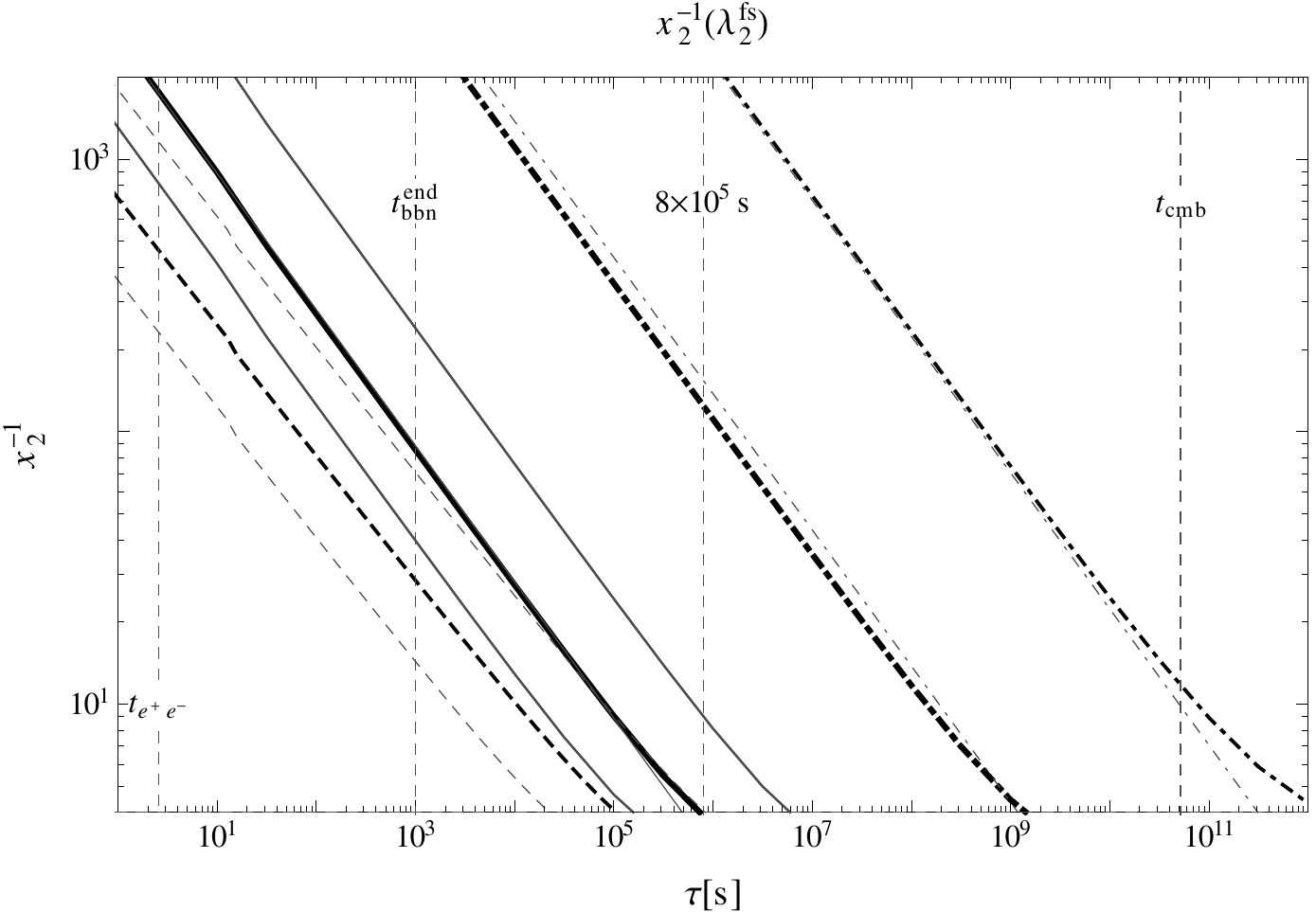}
\caption{The required $x_2=m_2/m$ depending on the time of decay to obtain a certain free-streaming scale $\l_2^\text{fs}$.
The thick solid curve corresponds to the numerical solution of~\eqref{ximplicit} for $\l_2^\text{fs}=0.4 \Mpc$. The area with $0.2 \Mpc \leq \l_2^\text{fs} \leq 1 \Mpc$ is found within the thin solid curves.
Values above the upper thin curve are excluded if the heavier daughter is to form the observed dark matter.
Values below the lower thin curve do not leave an observable imprint in the sky.
The heavier daughters form cold dark matter in this area, if $B_2$ satisfies~\eqref{nondomb2x2} at the boundary.
The dashed curves provide the same information for the analytic estimate~\eqref{x2desiredapprox}.
Above the thick dash-dotted curve $\l_2^\text{fs} > 10 \Mpc$. The thin dash-dotted curve corresponds to $\l_2^\text{fs} = 100 \Mpc$ and is given for better orientation.
Overplotted as very thin grey curves is the simple analytic approximation~\eqref{x2desiredfit}. Barely visible for $\l_2^\text{fs} = 0.4 \Mpc$ it is slightly too small for $10 \Mpc$, while the fit for $100 \Mpc$ is almost perfect again. 
We can read off that in a decay after $8\times 10^5 \seconds$ the decay products have to be emitted non-relativistically to obtain $\l_\text{fs}=0.4 \Mpc$, because the required $x_2$ becomes smaller than $1/4$. Other times are highlighted as in Fig.~\ref{fig:delta-tau-dr}.
}
\label{fig:x2desired}
\end{figure}
In principle, after being emitted also the heavier decay products may act as dark radiation throughout the history of the Universe.
However, they do not need to act as radiation at any time. 
At the boundary of \eqref{nondomb2x2approx}, the heavier daughters form the observed dark matter,
$\O_2 = \O_\text{dm}$, whose energy density depends on the energy density of dark radiation as explicated in Sec.~\ref{sec:structureformation}.
We find that they can, indeed, form the observed dark matter.
We will see how structure formation provides additional constraints on $B_2$ and $x_2$.
To act as cold or warm dark matter, for example, $x_2^{-1}$ must lie
below the upper thin solid curve in Fig.~\ref{fig:x2desired}.
If this bound is violated, tighter constraints than~\eqref{nondomb2x2} apply. For example, if they act as hot dark matter, we must replace $\O_\text{dm} \rightarrow \O_\text{hdm}^\text{max}$.
Since the free-streaming scale of the heavier daughters is set and can be adjusted by the involved couplings and masses,
there are not only constraints but also opportunities arising from the heavier daughters in structure formation.

\subsection{Solution to the missing satellites problem}
In simulated cold dark matter halos there is an overabundance of
substructures with respect to the observed number of Milky Way
satellites~\cite{Klypin:1999uc,Moore:1999nt}, which is known as the missing satellites problem.
The dispersion of structure on these small scales reduces the predicted number of galactic satellites.  
In particular, warm dark matter with a free-streaming scale
$\l_\text{fs} \gtrsim 0.2 \Mpc$ resolves the missing satellites
problem~\cite{Colin:2000dn,Lin:2000qq,Cembranos:2005us,Polisensky:2010rw}. We mentioned above the upper bound $\l_\text{fs} \lesssim 1 \Mpc$ from Lyman-$\a$ forest data.
In~\cite{Borzumati:2008zz} the authors considered the late decay of a massive particle with only one decay mode providing dark matter with the desired free-streaming scale. They found much smaller $\d$s than required by the consistent production of dark radiation, cf.~Sec.~\ref{sec:oddms}. Their numerical results agree with those obtained from our formulae~\eqref{lfsdeltatau} and~\eqref{lfsdeltatauafter}.
Neutralino dark matter from decays before BBN can have an appropriate
$\lambda_\text{fs}$ for large $\delta$ if the neutralinos lose enough
energy via scatterings with the thermal bath \cite{Hisano:2000dz}.  As
mentioned, we consider collisionless daughter particles, which is always
justified for decays after BBN.

We use the general result~\eqref{lfsintegrated} for the free-streaming scale of a particle emitted with relativistic momentum.
Repeating the steps leading from~\eqref{lfsintegrated} to~\eqref{lfsdeltatau}  with the only difference in
$\Tnr$, which is given here by~\eqref{Tnr12}, the free-streaming scale
$\l_2^\text{fs}$ of the heavier daughters as a function of $x_2$ and the time of decay is just~\eqref{lfsdeltatau} with the replacement
$
\frac{(\d+1)^2-1}{\d+1} \rightarrow (x_2^{-2}-4)^\frac{1}{2} 
$.
This can be read as an implicit equation for $x_2$ arising from constraints on its free-streaming scale or to obtain a desired free-streaming scale to solve the missing satellites problem, for example. It is
\begin{eqnarray}
\label{ximplicit}
 \frac{\l_2^\text{fs}(x_2,\t)}{0.4 \Mpc} &\approx& 0.23
\left(\frac{\t}{10^7 \seconds}\right)^\frac{1}{2}
(x_2^{-2}-4)^\frac{1}{2} 
\frac{(\gastd)^{1/4}}{(\gastnr)^{1/2}}
\left(\frac{(\gastsnr)^2}{\gastsd \gastst}\right)^\frac{1}{3}  \nonumber \\
& \times & \Bigg\lbrace  
5
- \frac{4}{\mu} (x_2^{-2}-4)^{-\frac{1}{2}} 
\left(\frac{\gastnr}{\gastd}\right)^\frac{1}{4}
\left(\frac{\gastsd}{\gastsnr}\right)^\frac{1}{3}  \nonumber \\
 &+& \ln \left\lbrack 
\frac{\teq}{\t} \frac{4}{\m^2} 
(x_2^{-2}-4)^{-1} 
\left(\frac{\gastnr}{\gastd}\right)^\frac{1}{2}
\left(\frac{\gastsd}{\gastsnr}\right)^\frac{2}{3}
\right\rbrack 
\Bigg\rbrace \, .
\end{eqnarray}
To obtain the thick solid curve in Fig.~\ref{fig:x2desired} we solved~\eqref{ximplicit} numerically.
Equally well, one can read off the required time of decay from Fig.~\ref{fig:x2desired}, if $x_2$ is set by the particle physics model.
Overplotted as very thin grey curves is the simple analytic approximation
\begin{equation}
\label{x2desiredfit}
 x_2 \simeq 0.1
\left(\frac{0.4 \Mpc}{\l_2^\text{fs}} \right)^{1.21}
\left(\frac{\t}{10^5 \seconds}\right)^{0.5} .
\end{equation}
For $\l_2^\text{fs}= 0.4 \Mpc$ it is barely visible except very close to $8 \times 10^5 \seconds$. For $\l_2^\text{fs}= 10 \Mpc$ it is slightly too small, while for $\l_2^\text{fs}= 100 \Mpc$ the fit is again almost perfect.

In order to find an analytic estimate we first simplify~\eqref{ximplicit} by omitting relativistic free-streaming and free-streaming after $\teq$. 
Exploiting logarithm rules we can see that one might neglect the factor $(x_2^{-2}-4)^{-1}$ in the logarithm, if
$
x_2 \gg (\m/2) (\t/\teq)^{1/2} (\gastd/\gastnr)^{1/4} (\gastsnr/\gastsd)^{1/3}
$,
which holds for decays sufficiently earlier than $\teq$.
Then $x_2$ can be singled out to obtain
\begin{eqnarray}
\label{x2desiredapprox}
 x_2 &\simeq& \Bigg[\Big(0.23
\frac{0.4 \Mpc}{\l_2^\text{fs}}
\left(\frac{\t}{10^7 \seconds}\right)^\frac{1}{2}
\frac{(\gastd)^\frac{1}{4}}{(\gastsnr)^\frac{1}{2}}
\left(\frac{(\gastsnr)^2}{\gastst \gastsd}\right)^\frac{1}{3} \nonumber \\
&\times &
\ln \left\lbrack
\frac{\teq}{\t}
\frac{4}{\m}
\left(\frac{\gastnr}{\gastd}\right)^\frac{1}{2}
 \left(\frac{\gastsd}{\gastsnr}\right)^\frac{2}{3}
\right\rbrack
\Big)^{-2} +4 \Bigg]^{-\frac{1}{2}} \, ,
\end{eqnarray}
where $\gastnr=\gastt$ and $\gastsnr=\gastst$ for $\l_2^\text{fs} = 0.4 \Mpc$.
As can be seen in Fig.~\ref{fig:x2desired} the result is systematically too large and the logarithmic dependence is misleading. On the other hand we find the dependence on $\t$ and a very weak dependence on $\gastd$. The shift with $\l_2^\text{fs}$ is well reproduced. 

The corresponding time when the heavier daughters become
non-relativistic is about $3.4 \times 10^6 \seconds$. So this event appears rather unobservable.
It is not a surprise that we find a constant value, because in this case $\l_\text{fs}= 0.4 \Mpc$ is set by $t_2^\text{nr}$ as argued below~\eqref{lfsintegrated}.
With the analytic result~\eqref{x2desiredapprox} we can verify the dependence of $t_2^\text{nr}$ on $\gastd$. It is weaker than $\propto (\gastd)^{1/6}$ and mainly due to the change in the contribution from relativistic free-streaming.

Inserting the desired $x_2$ into the non-dominance
requirement~\eqref{nondomb2x2} at the boundary,
where the heavier daughters actually
form the observed dark matter, yields the unique branching ratio $B_2$
required to produce the desired dark radiation and the dark matter of the Universe with the desired free-streaming scale at the same time from the same late-decaying particle.
We find a constant value,
\begin{equation}
\label{B2desiredconst}
B_2 \simeq 5.6 \times 10^{-3}   
\left(\frac{\l_2^\text{fs}}{0.4 \Mpc}\right) \D\Neff^{-1}
\left(\frac{\omegadm}{0.1286}\right) ,
\end{equation}
with an even weaker dependence on $\gastd$ than $t_2^\text{nr}$. This is no
longer a surprise, because with fixed $t_2^\text{nr}$ also $B_2$ becomes fixed in order for the heavier daughters to form the observed dark matter.
Taking into account uncertainties, $8.1 \times 10^{-4} \lesssim B_2 \lesssim 2.8 \times 10^{-2}$.
In the course of the calculation we had to discard the second solution to~\eqref{ximplicit} as unphysical.
To obtain an analytic estimate we insert~\eqref{x2desiredapprox} into~\eqref{nondomb2x2} and single out
\begin{equation}
 \label{b2analytic}
B_2 \simeq \frac{1}{B_2^{-1} -1} = 0.040
\left(\frac{\l_2^\text{fs}}{0.4 \Mpc}\right) \D\Neff^{-1}
\left(\frac{\omegadm}{0.1286}\right) 
\ln \left[
\frac{\teq}{\t}
\frac{4}{\m^2}
\left(\frac{\gastnr}{\gastd}\right)^\frac{1}{2}
 \left(\frac{\gastsd}{\gastsnr}\right)^\frac{2}{3}
\right]^{-1} ,
\end{equation}
where we took $1-4 x_1^2 \simeq 1$ assuming sufficiently small $x_1$ and, again, $\gastnr=\gastt$ and $\gastsnr=\gastst$.
The logarithmic dependence is misleading, while the dependencies on $\l_2^\text{fs}$, $\D\Neff$ and $\omegadm$ in~\eqref{B2desiredconst} find a reason.

\subsection{Hot dark matter opportunity}
 
In this section we point out the same hot dark matter (HDM) opportunity as at the end of Sec.~\ref{sec:hdm} but for the case of two dark decay modes.
To be able to lead to a non-zero cosmological neutrino mass scale like $\sum m_\n =0.34 \eV$ inferred from the cluster abundance in~\cite{Benson:2011ut} the free-streaming scale of the heavier daughters must be larger than the corresponding structure formation scale, $\l_2^\text{fs} > \l_\text{gc} \sim 10 \Mpc$.
As before we solved~\eqref{ximplicit} numerically to obtain the required mass hierarchy $x_2$ as depicted in Fig.~\ref{fig:x2desired} to have $\l_2^\text{fs} \simeq 10 \Mpc$. 
The corresponding times for the heavier daughters to become non-relativistic are $t_2^\text{nr}(\l_2^\text{fs}=10 \Mpc)\simeq 2.4 \times 10^9 \seconds$ and  $t_2^\text{nr}(\l_2^\text{fs}=100 \Mpc)\simeq 8.6 \times 10^{11} \seconds$.
Thus only for very large free-streaming scales they become non-relativistic after $\taumax$.
Inserting the found $x_2$ into~\eqref{nondomb2x2} with the replacement
$\Omega_\text{dm} h^2 \rightarrow \Omega_\text{hdm} h^2 \simeq \sum m_\nu / 93\eV$,
we find at the boundary the branching ratio into the heavier daughters
$B_2$ that yields $\D\Neff=0.86$ and $\sum m_\n =0.34 \eV$ from the same
decaying particle.
In order to act as the dark radiation we assume the lighter daughters to be still relativistic today with the corresponding mass hierarchy $x_1$ given by~\eqref{x12}.
We find a constant branching ratio
\begin{equation}
 \label{B2hdm}
B_2 \simeq 9 \times 10^{-3}
\left(\frac{\l_2^\text{fs}}{10 \Mpc}\right)
\left(\frac{\Omega_\text{hdm}h^2}{0.0037}\right)
\D\Neff^{-1} \,.
\end{equation}
Larger free-streaming scales are possible or might be preferred.
At $\l_2^\text{fs}= 100\Mpc$ the approximation~\eqref{B2hdm} is significantly smaller than the true value $0.18$.
In any case, it is trivial to solve~\eqref{nondomb2x2} with the appropriate replacements for any fixed value of $x_2$.
The allowed area of values $x_2^{-1}(\t)$ is found above the thick dash-dotted curve in Fig.~\ref{fig:x2desired}.

Altogether, we have shown that any desired amount of HDM and dark
radiation can originate from the decay of the same particle, while in
the case of two dark decay modes the relative branching ratio allows to
vary the time the HDM becomes non-relativistic. This time determines the
HDM free-streaming scale.
In the case of massive neutrinos and also in the case of HDM from
particle decay with only one dark decay mode, this time is always after
$\teq$. This offers a possibility to distinguish between these
cases in cosmological observations.

\section{Results and conclusions}
\label{sec:conclusions}
We studied particle decay as the origin of dark radiation.
After elaborating general properties of such cosmologies we determined model-independent constraints
on possible underlying theories.
Since the energy density of the decaying particle is fixed by the amount of dark radiation and thus by observations independent of an underlying particle physics model, bounds on branching ratios and constraints on the mass hierarchies between decaying particle and decay products depend in a unique way on the time of decay.

If the decaying particle possesses only one dark decay mode, we find that the minimal free-streaming scale of its decay products is so large that hot dark matter constraints apply to their relic densities. 
Therefore, the heavier decay product in a particle decay producing the desired dark radiation cannot form the observed dark matter.
So hot dark matter constraints determine the minimal mass hierarchy between decaying particle and the heavier decay product. This constraint is depicted in Fig.~\ref{fig:delta-tau-dr}, where also cosmology-specific uncertainties are shown.
The hot dark matter bound is the tightest bound obtainable taking into account the impact of the scenario on structure formation.
On the other hand, a hot dark matter component in excess of the SM neutrinos could not only have originated from particle decay but also share its origin with the desired dark radiation. The heavier decay product can form a finite hot dark matter component, while the lighter one acts as dark radiation.
In any case decay products become non-relativistic during or after CMB times.

If the decaying particle possesses two dark decay modes, the free-streaming scales of the decay products are set and can be adjusted by the involved couplings and masses. Therefore, there are not only constraints, as shown in Fig.~\ref{fig:xmin}, but also additional opportunities arising from the impact of the heavier decay products on structure formation.
Depending on the time of decay we provide the unique mass hierarchy and relative branching to produce dark radiation and dark matter with any desired free-streaming scale  from the same particle decay. 
The observed dark matter satisfying the cold dark matter paradigm may have originated from such a decay. Any finite hot dark matter contribution can be explained and possibly be distinguished in future cosmological observations from other sources like SM neutrinos.
In a different range of mass hierarchies and the corresponding branching ratios, the dark matter from particle decay solves the missing satellites problem.

We determined general upper bounds on several branching ratios of the decaying particle into decay products with SM interactions. These are independent of the underlying theory and, for example, independent of the number of dark decay modes.
Direct decays are constrained as shown in Figs.~\ref{fig:BemmaxBBNdirect},~\ref{fig:BhadmaxBBNdirect} and~\ref{fig:BphmaxCMBdirect}. 
Since these constraints are severe, we considered scenarios fulfilling them by construction.
Dangerous terms in the Lagrangian are forbidden by kinematics or by symmetries. 
However, after taking into account the amount of energy that is always carried away invisibly we find that also
off-shell and loop processes are severely constrained, cf.~Figs.~\ref{fig:BemmaxBBN},~\ref{fig:BhadmaxBBN} and~\ref{fig:BphmaxCMB}.
A certain finite branching ratio into hadronically interacting particles could solve the cosmic lithium problems.
The emission of photons may enable future CMB polarimeters to detect and identify the desired cosmological particle decay.
More robustly, the obtained bounds have the power to exclude many particle physics scenarios.
As an example we show how decays of the lightest ordinary supersymmetric particle into an invisible particle like the gravitino are excluded as the origin of the desired dark radiation during and after BBN\@.
We argue that particle decay as origin of dark radiation serves as a motivation to extend existing studies of constraints on cosmological particle decays towards smaller masses of the decaying particle.
For example, BBN constraints do not seem to apply to a particle that freezes out relativistically and produces the desired dark radiation in its necessarily late decay during or after BBN.

Since we provide simple analytic formulae and figures pointing out uncertainties, our results
can easily be adopted to constrain particle physics models and as a guideline for model building.
Particle decay as the origin of dark radiation raises very specific requirements on any underlying theory.
Most existing proposals assume implicitly that constraints are satisfied. Often decaying  particle and decay products have only extremely weak interactions. Thus, they safely satisfy branching ratio constraints but cannot lead to additional observable consequences.

We point out various opportunities of a cosmological particle decay serving as a motivation for further studies.
There is a plethora of possible cosmologies to be explored.
As close as some interplay between the different mysteries of our universe is, as attractive appears a theory that combines and intertwines them.

\subsection*{Acknowledgements}
We thank Steen Hannestad, Jan Hamann, Raul Jimenez and Licia Verde for valuable discussions.
We especially thank Torsten Bringmann for sharing his expertise.
This work was supported by the German Research Foundation (DFG) via the
Junior Research Group ``SUSY Phenomenology'' within the Collaborative
Research Centre 676 ``Particles, Strings and the Early Universe''.

\begin{appendix}

\section{Exponential decay in an expanding universe}
\label{appendix:expdecay}
Any initial momentum $p_\text{ini}$ from a decay is red-shifted by the expansion of the Universe. It is
\begin{equation}
\label{poft}
 p(t,t_0)=p_\text{ini} \frac{a(t)}{a(t_0)} = p_\text{ini} \left(\frac{t}{t_0}\right)^\frac{2}{3(1+\o)} \, .
\end{equation}
How the scale factor $a \propto t^\frac{2}{3(1+\o)}$ grows with time depends on the equation of state of the Universe,
$p=\o \r$ with $\o=1/3$ if it is radiation-dominated and $\o=0$ in the case of matter domination.
Earlier emitted particles experience a longer time of expansion and thus more red-shift than later emitted particles. Taking into account the exponential decay law this leads to a more involved momentum distribution  than the monochromatic line obtained in the sudden decay approximation.

Two-body decays have no intrinsic momentum distribution.
In this case the momentum distribution function $f(p,t_0)$ of an emitted particle is determined by the number of produced particles in a given time interval $dt$ at time $t$. It is
\begin{equation}
 N \frac{dt}{\t} =  f(p,t_0) dp \, ,
\end{equation}
 if $N(t) =N_0 e^{-t/\t}$ is the number of decaying particles. If the number of dark radiation particles produced in each decay $\gdr \neq 1$, we replace $N \rightarrow \gdr N$.
So
\begin{equation}
 f(p,t_0) = \frac{N}{\t}\frac{dt}{dp}
\end{equation}
and reversing~\eqref{poft} we can insert $t(p,t_0)$, which gives $dt/dp$ as well, also in $N(t)$ arriving at
\begin{eqnarray}
\label{fofpt0}
 f(p,t_0) &=& c N_0 p^{-c}(\t,t_0) e^{-\left(\frac{p}{p(\t,t_0)}\right)^c} p^{c-1} \nonumber \\
 &=& c N_0 p^{-1} \frac{t_0}{\tau} \left(\frac{p}{p_\text{ini}}\right)^c e^{- \frac{t_0}{\tau} \left(\frac{p}{p_\text{ini}}\right)^c}    \, ,
\end{eqnarray}
where we abbreviated $c\equiv 3(1+\o)/2 > 0$.%
\footnote{ This distribution agrees with the one obtained in the appendix of \cite{Scherrer:1987rr}.}
With the assumption of a constant $\o$ we do not include the case of a particle that shortly dominates the energy density of the Universe at its decay during radiation domination. As argued in the text this situation is not expected to occur.
\begin{figure}[t]
 \centering
   \includegraphics[width=0.9 \textwidth]{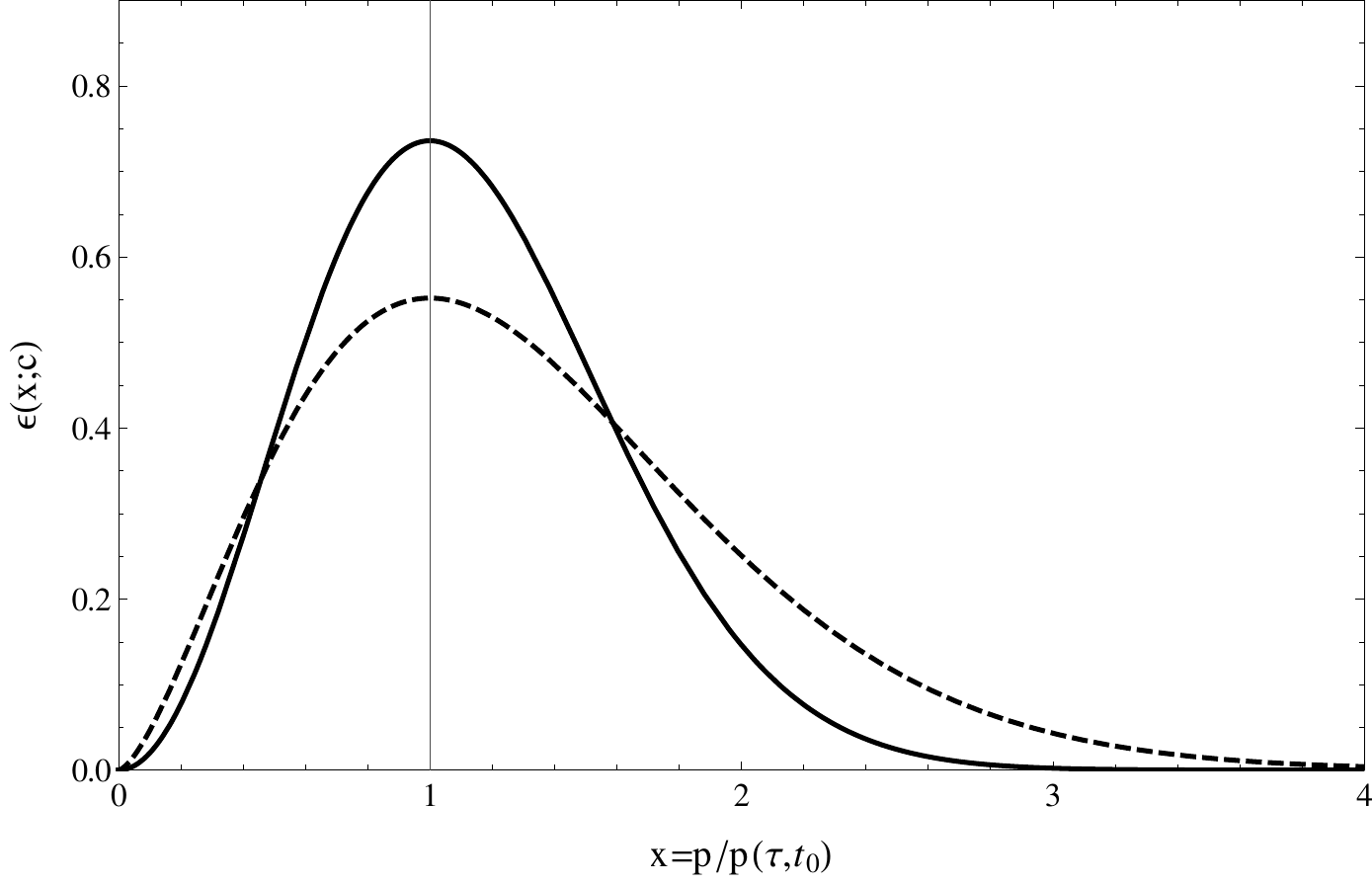}
\caption[$\e(x)$]{Normalised, time-invariant energy spectrum $\e(x=p/p(\t,t_0);c)$ of relativistic particles from two-body decay (decaying particle at rest) in an expanding Universe with scale factor $a\propto t^{1/c}$.
The solid (dashed) line is obtained from~\eqref{energyspectrum} as described in the text with $c=2 (3/2)$.
The maximum of the energy spectra is highlighted at $x=1 \Leftrightarrow p=p(\t,t_0)$.
}
\label{fig:energyspec}
 \end{figure}
\begin{figure}[t]
 \centering
   \includegraphics[width=0.9 \textwidth]{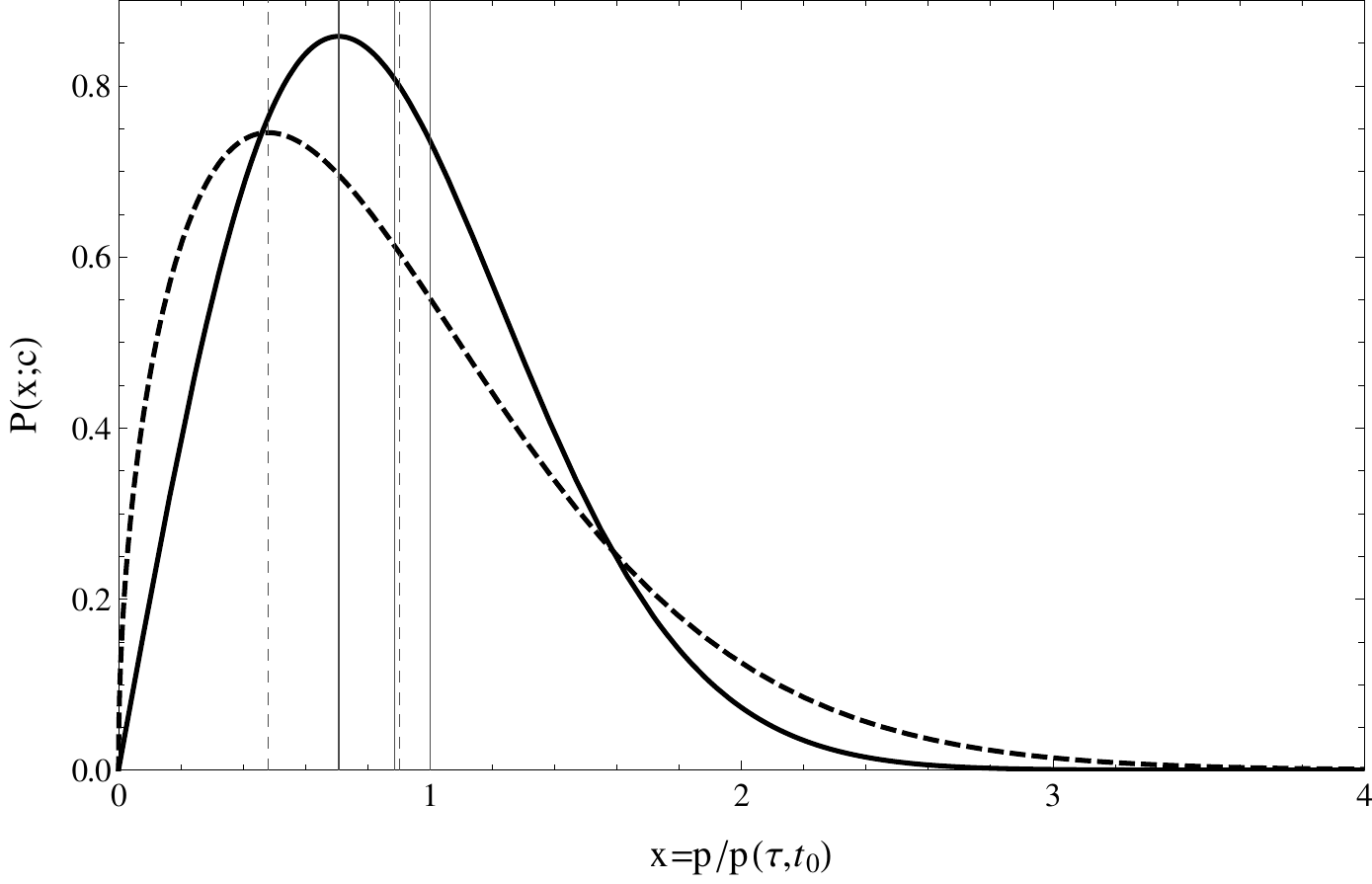}
\caption[$P(x)$]{
Normalised, time-invariant probability distribution $P(x;c)=c x^{c-1} e^{-x^c}$ of~\eqref{probdistr} for finding a relativistic particle from two-body decay  (decaying particle at rest) within the infinitesimal momentum interval $[p,p+\d p]$.
The solid (dashed) line is obtained for $c=2 (3/2)$.
Vertical lines highlight the maxima of the corresponding distributions, $x\simeq 0.707$ ($c=2$) and $x\simeq0.481$ ($c=3/2$), as well as the mean of the distributions, $\m = \sqrt{\pi}/2$ ($c=2$) and $ \m = \G[2/3] 2/3$ ($c=3/2$).
The maximum of the corresponding energy spectra is at $x=1$, cf.~Fig.~\ref{fig:energyspec}. 
}
\label{fig:probdistr}
\end{figure}

\paragraph{Total energy}
Since dark radiation particles are relativistic, each particle's energy is given by its kinetic energy, so its momentum, $E \simeq p$. The total energy in dark radiation is thus given by the integral
\begin{equation}
\label{Etot}
 E_\text{dr}(t_0) =\int\limits_0^{p_\text{ini}} p f(p,t_0)   dp \, .
\end{equation}
Inserting~\eqref{fofpt0} for $c>0$ this becomes
\begin{equation}
\label{Edrgen}
 E_\text{dr}(t_0)= N_0  p(\t, t_0) \frac{1}{c} \left( \G[\frac{1}{c}] - c \G[1 + \frac{1}{c}, \frac{t_0}{\t} ] \right) \, ,
\end{equation}
where $\G[z]$ denotes the Euler gamma function and $\G[a,z]$ the incomplete gamma function.
In the case of radiation domination ($c=2$)~\eqref{Edrgen} reduces to
\begin{equation}
 E_\text{dr}(t_0)= N_0  p(\t, t_0) \left( \frac{\sqrt{\pi}}{2} \text{Erf}[\left(\frac{t_0}{\t}\right)^\frac{1}{2}] - \left(\frac{t_0}{\t}\right)^\frac{1}{2} e^{-\frac{t_0}{\t}} \right) \, ,
\end{equation}
where Erf$[z]$ denotes the error function.

Times of interest are later than the time of decay, so that we should investigate the limit $t_0 \gg \t$.
In this limit the general expression~\eqref{Edrgen} yields
\begin{equation}
\label{Edrlim}
 \lim_{t_0\gg \t} E_\text{dr}(t_0) = N_0 p(\t,t_0) \frac{1}{c} \G[\frac{1}{c}]  \, .
\end{equation}
The difference by taking into account the exponential decay behaviour in contrast to the sudden decay approximation is thus found in the factor $c^{-1} \G[c^{-1}]$. For the two important cases this is
\begin{equation}
 \mu =\frac{\sqrt{\pi}}{2} \simeq 0.886 \qquad \text{in radiation domination, } c=2,
\end{equation}
and
\begin{equation}
 \mu = \frac{2}{3} \G[\frac{2}{3}] \simeq 0.902 \qquad \text{in matter domination, } c=3/2 \, .
\end{equation}
The difference in~\eqref{Edrlim} between these cases is thus below two percent.
However, compared to the sudden decay approximation it is twelve percent.

\paragraph{Invariant energy spectrum}
The integral of the momentum distribution function $f(p,t_0)$ over the full parameter space yields the total number of particles $g_\text{dr} N_0$.
The integral of the energy spectrum $\e$ over full parameter space yields the total energy~\eqref{Etot} and is thus just found as 
\begin{equation}
\label{energyspectrum}
\e(p,t_0) = p f(p,t_0)\, .
\end{equation}

The effect of red-shift is easily comprehensible. At times $t_0 \gg \t$
it is an equal shift of all momenta towards smaller ones, which does not
change the characteristic form developed roughly until $t_0 \sim 3 \t$
and $t_0 \sim 4 \t$ for radiation and matter domination, respectively. Therefore, we give the normalised and time-invariant energy spectrum $\e(x)$ in Fig.~\ref{fig:energyspec} by defining $x=p/p(\t,t_0)$. 

We treat the non-relativistic decaying particle to be at rest. The correction from a non-zero kinetic energy is negligible, if the decaying particle has a sufficiently small momentum. This is the case for all considered scenarios in this work.

\paragraph{Probability distribution}
Determining the probability distribution $P(p,t_0)$ for finding a particle within an infinitesimal momentum interval $[p,p+\d p]$ we note that by construction
\begin{equation}
 1 =  \int_0^\infty  P(p,t_0)dp = \int_0^\infty \frac{f(p,t_0)}{N_0} dp \, .
\end{equation}
Thus we have also just found $P(p,t_0)$. Performing the same change of coordinates as for the energy spectrum, $p \rightarrow x=p/p(\t,t_0)$, we obtain
\begin{equation}
\label{probdistr}
 \int_0^\infty P(x) dx =\int_0^\infty c x^{c-1} e^{-x^c}dx  \,.  
\end{equation}
This time-invariant distribution is depicted in Fig.~\ref{fig:probdistr}. 
Highlighted are the maximum of the corresponding energy spectrum at $x=1$, which is independent of $c$,
 the maximum of the distribution itself given by $(1-1/c)^{1/c}$ at $x\simeq 0.707$ ($c=2$) or $x\simeq 0.481$ ($c=3/2$), and the mean of the distribution 
\begin{equation}
\mu(P) = \int_0^\infty  x P(x)dx  =c^{-1}\G[c^{-1}] \,,
\end{equation}
to be compared with~\eqref{Edrlim}.
Obviously, $P(x)$ is asymmetric.
The variance $\phi(P)$, also known as the second central moment of $P$, is found as
\begin{equation}
\phi(P)=\int_0^\infty  (x-\mu)^2 P(x)dx
 = \G[1+\frac{2}{c}] - \mu^2 \simeq \begin{cases}
            0.215 & \text{for } c=2 \\
   0.376 & \text{for } c=3/2
          \end{cases} \,.
\end{equation}
The difference in $\phi$ is thus about $55\%$. In the sudden decay approximation the variance
vanishes by definition. For the same reason the root-mean-square velocity, also known as the second moment about zero, equals the mean velocity.
We have just found the correction factor for the root-mean-square velocity $\phi_\text{rms} = \sqrt{\G[1+\frac{2}{c}]}$.
It is one for $c=2$ and about $1.09$ for $c=3/2$.
\end{appendix}

\phantomsection 
\addcontentsline{toc}{chapter}{References}
\bibliography{StructForm}

\end{document}